\documentclass[11pt]{article}

\usepackage{ntabbing}
\usepackage{amssymb}
\usepackage{graphicx}
\usepackage{color}
\usepackage{enumerate}
\usepackage{algorithm}
\usepackage{a4wide}

\def\comm#1{}
\def\incomm#1{}

\newcommand{\bs}{\backslash}

\def\PF{\noindent{\em Proof:} }

\def\QED{\mbox{}\hspace*{\fill}{$\Box$}\medskip}

\newtheorem{thm}{Theorem}
\newtheorem{lem}[thm]{Lemma}
\newtheorem{propo}[thm]{Proposition}
\newtheorem{defi}[thm]{Definition}
\newtheorem{corol}[thm]{Corollary}

\newtheorem{repeatthm}{Theorem}

\newcommand{\RIS}{\mu} 
\newcommand{\freedom}{\lambda}

\newcommand{\TAR}{\mbox{TAR}}
\newcommand{\TJ}{\mbox{TJ}}
\newcommand{\tar}{\leftrightarrow} 
\newcommand{\tj}{\leftrightarrow_{\mbox{\footnotesize TJ}}}
\newcommand{\TARseq}{TAR-sequence}
\newcommand{\TJseq}{TJ-sequence}
\newcommand{\indset}{independent set}
\newcommand{\Mc}{\overline{M}}

\newcommand{\GG}{\mathcal{G}}

\begin{document}

\title{Independent Set Reconfiguration in Cographs}
\date{\today}
\author{Paul Bonsma\footnote{University of Twente, Faculty of EEMCS, PO Box 217, 7500 AE Enschede, the Netherlands.}}

\maketitle

\begin{abstract}
We study the following independent set reconfiguration problem, called {\em \TAR-Reachabi\-lity:} given two \indset s $I$ and $J$ of a graph $G$, both of size at least $k$, is it possible to transform $I$ into $J$ by adding and removing vertices one-by-one, while maintaining an \indset\ of size at least $k$ throughout? This problem is known to be PSPACE-hard in general. For the case that $G$ is a cograph (i.e. $P_4$-free graph) on $n$ vertices, we show that it can be solved in time $O(n^2)$, and that the length of a shortest reconfiguration sequence from $I$ to $J$ is bounded by $4n-2k$, if such a sequence exists.

More generally, we show that if $\GG$ is a graph class for which (i) \TAR-Reachability can be solved efficiently, (ii) maximum \indset s can be computed efficiently, and which satisfies a certain additional property, then the problem can be solved efficiently for any graph that can be obtained from a collection of graphs in $\GG$ using disjoint union and complete join operations. Chordal graphs are given as an example of such a class $\GG$.
\end{abstract}

\section{Introduction}
\label{sec:intro}

Reconfiguration problems have been studied often in recent years.
These arise in settings where the goal is to transform feasible solutions to a problem in a step-by-step manner, while maintaining a feasible solution throughout.
A {\em reconfiguration problem} is obtained by defining {\em feasible solutions} (or configurations) for {\em instances} of the problem, and a (symmetric) {\em adjacency relation} between solutions. This defines a {\em solution graph} for every instance, which is usually exponentially large in the input size. Usually, it is assumed that {\em adjacency} and {\em being a feasible solution} can be tested in polynomial time.
Typical questions that are studied are deciding the existence of a path between two given solutions {\em (reachability)}, finding shortest paths between solutions, deciding whether the solution graph is connected or giving sufficient conditions for this, and giving bounds on its diameter.
For example, the literature contains such results on the reconfiguration of vertex colorings~\cite{BC09,CHJ08,CHJ09,CHJ11}, boolean assignments that satisfy a given formula~\cite{GKM09}, independent sets~\cite{HD05,IDH11,KMM12,MNRSS13}, matchings~\cite{IDH11}, shortest paths~\cite{PSPR_FSTTCS12,TCS13,KMM11TCS}, subsets of a (multi-)set of integers~\cite{EW12,ID11}, etc. Techniques for many different reconfiguration problems are discussed in~\cite{IDH11,MNRSS13}. See the recent survey by van den Heuvel~\cite{JvdH13} for an overview of and introduction to reconfiguration problems, and a discussion of their various applications.

One of the most well-studied problems of this kind is the reconfiguration of {\em independent sets}. For a graph $G$ and integer $k$, the independent sets of size at least/exactly $k$ of $G$ form the feasible solutions. Independent sets are also called {\em token configurations}, where the independent set vertices are viewed as {\em tokens.} Three types of adjacency relations have been studied in the literature: in the {\em token jumping (TJ)} model~\cite{IDH11}, a token can be moved from any vertex to any other vertex. 
In the {\em token sliding (TS)} model, tokens can be moved along edges of the graph~\cite{HD05}. In the {\em token addition and removel (TAR)} model~\cite{IDH11}, tokens can be removed and added in arbitrary order, though at least $k$ tokens should remain at any time ($k$ is the {\em token lower bound}). 
Of course, in all of these cases, an independent set should be maintained, so tokens can only be moved/added to vertices that are not dominated by the current token configuration. 

The {\em reachability problem} has received the most attention in this context: given two independent sets $I$ and $J$ of a graph $G$, and possibly a token lower bound $k\le \min\{|I|,|J|\}$, is there a path (or {\em reconfiguration sequence}) from $I$ to $J$ in the solution graph? We call this problem {\em TJ-Reachability, TS-Reachability} or {\em TAR-Reachability}, depending on the adjacency relation that is used.
Kami\'{n}ski et al~\cite{KMM12} showed that the TAR-Reachability problem generalizes the TJ-Reachability problem (see Section~\ref{sec:prelim} for details).
For all three adjacency relations, this problem is PSPACE-hard, even in perfect graphs~\cite{KMM12}, and even in planar graphs of maximum degree 3~\cite{HD05}. (The latter result is not explicitly stated in~\cite{HD05}, but can easily be deduced from the given reduction. 
See~\cite{BC09} for more information.) See also~\cite{IDH11} for an alternative, simple PSPACE-hardness proof. In addition, in~\cite{KMM12}, the problem of deciding whether there exists a path of length at most $l$ between two solutions is shown to be strongly NP-hard, for all three adjacency models.

On the positive side, these problems can be solved in polynomial time for various restricted graph classes. 
The result on matching reconfiguration by Ito et al~\cite{IDH11} implies that for line graphs, TJ-Reachability and TAR-Reachability can be solved efficiently. In~\cite{KMM12}, an efficient algorithm is given for TS-Reachability in cographs, and it is shown that for TJ-Reachability in even-hole-free graphs, a reconfiguration sequence exists between any pair of \indset s $I$ and $J$, and that the shortest reconfiguration sequence always has length $|I\bs J|$.

\paragraph{New results and techniques}
In this paper, we show that TAR-Reachability and TJ-Reachability can be solved in time $O(n^2)$ for cographs, where $n$ is the number of vertices of the input graph. This answers an open question from~\cite{KMM12}. In addition, we show that for cographs, components of the solution graph have diameter at most $4n-2k$ and $2n-k$, under the TAR-model and TJ-model, respectively. Recall that a graph is a {\em cograph} iff it has no induced path on four vertices. Alternatively, cographs can be defined as graphs that can be obtained from a collection of trivial (one vertex) graphs by repeatedly applying {\em (disjoint) union} and {\em (complete) join} operations. The order of these operations can be described using a rooted {\em cotree}. 
This characterization allows efficient dynamic programming (DP) algorithms for various NP-hard problems. Our algorithm is also a DP algorithm over the cotree, albeit more complex than many known DP algorithms on cographs. 
For both solutions $A$ and $B$, certain values are computed, using first a {\em bottom up} DP phase, and next a {\em top down} DP phase over the cotree. Using these values, we can conclude whether $B$ is reachable from $A$. 
Because of this method, we in fact obtain a stronger result: TJ- and TAR-Reachability can be decided efficiently for any graph that can be obtained using join and union operations, when starting with a collection of base graphs from a graph class $\GG$ that satisfies the following properties:
\begin{itemize}
\item For any graph in $\GG$, the TAR-Reachability problem can be decided efficiently, and
\item for any graph in $\GG$ and \indset\ $I$, the size of a maximum independent set that is TAR-reachable from $I$ can be computed efficiently, for all token lower bounds $k\le |I|$.
\end{itemize}
In this paper, we show that an example of such a graph class is the class of {\em chordal graphs}. In another paper, we show that the class of {\em claw-free graphs} also satisfies these properties~\cite{BKW}. Combining these results yields quite a rich graph class for which this PSPACE-hard problem can be solved in polynomial time.

Another motivation for this research is that cographs form the base class for various graph width measures: cographs are exactly the graphs of cliquewidth at most two, and exactly the graphs of modular-width two~\cite{CO00}. The corresponding graph decompositions ($k$-expressions and modular decompositions) have been well-studied in algorithmic graph theory, because of the fact that many NP-hard problems can be solved efficiently on graphs where the width of these decompositions is low, using DP algorithms~\cite{CMR00,GLO13}. Another similar, successful and widely used notion is that of a tree decomposition / the treewidth of a graph~\cite{Bod93tourist}. 
The success of such approaches for NP-complete problems and NP-optimization problems is unmistakable in the area of algorithmic graph theory. However, surprisingly, no nontrivial results of this kind are known for reconfiguration problems, to our knowledge. 
More precisely: we are not aware of any reconfiguration problems that are PSPACE-hard in general, but that can be solved efficiently on graphs of treewidth or cliquewidth at most $k$, for every constant $k$. On the other hand, none of the studied reconfiguration problems have been shown to be PSPACE-hard on graphs of bounded treewidth/cliquewidth. We expect that positive results of this kind are certainly possible, but have not yet been obtained due to the lack of DP techniques for reconfiguration problems. This paper gives a first example of how dynamic programming over graph decompositions can be used successfully for PSPACE-hard reconfiguration problems. 
This is a first step towards solving various reconfiguration problems for graphs of bounded (modular-, clique-, tree-) width; we expect that similar algorithmic techniques can be used and are necessary to show that indeed, various reconfiguration problems can be solved efficiently using DP over graph decompositions.
We remark that a DP approach has also been used to show that the PSPACE-hard Shortest Path Reconfiguration problem can be solved in polynomial time on planar graphs~\cite{PSPR_FSTTCS12}, although a problem-specific layer decomposition of the graph was used.

Our DP algorithm for the TAR-Reachability problem is presented in Sections~\ref{sec:ModuleLemmas}--\ref{sec:summary}. 
First, in Section~\ref{sec:outline}, an example is given, the proof of this statement is outlined, and a detailed overview of Sections~\ref{sec:ModuleLemmas}--\ref{sec:summary} is given. In Section~\ref{sec:graphclasses}, examples of graph classes are given for which this algorithm works; in particular graphs obtained from chordal graphs using union and join operations (which includes cographs).
The bound on the diameter of the solution graph is given in Section~\ref{sec:diameter}. We start in Section~\ref{sec:prelim} with precise definitions, and end in Section~\ref{sec:discussion} with a discussion.

\section{Preliminaries}
\label{sec:prelim}

\paragraph{Token Addition and Removal}
By $\alpha(G)$ we denote the maximum size of an independent set in $G$.
In this paper, we use the {\em token addition and removal (TAR)} model for independent set reconfiguration. 
For a graph $G$ and integer $k$, the vertex set of the graph $\TAR_k(G)$ is the set of all independent sets of size at least $k$ in $G$. Two distinct independent sets $I$ and $J$ are adjacent in $\TAR_k(G)$ if there exists a vertex $v\in V(G)$ such that $I\cup \{v\}=J$ or $I=J\cup \{v\}$.
Vertices from \indset s will also be called {\em tokens}, and we will also say that $J$ is obtained from $I$ by {\em adding one token on $v$} resp.\ {\em removing one token from $v$}, or that {\em $J$ is obtained from $I$ using one \TAR-step.}

For an integer $k$ and two independent sets $I$ and $J$ of $G$ with $|I|\ge k$ and $|J|\ge k$, we write $I\tar_k^G J$ if $\TAR_k(G)$ contains a path from $I$ to $J$.
Observe that $I\tar_0^G J$ always holds, and that the relation $\tar_k^G$ is an equivalence relation, for all $G$ and $k$.
The superscript $G$ is omitted if the graph in question is clear.
If $G$ and $k$ are clear from the context, we will also simply say that {\em $J$ is reachable from $I$}.
A sequence $I_0,\ldots,I_k$ is called a {\em $k$-\TARseq}\ for $G$ from $I_0$ to $I_k$ if
\begin{itemize}
 \item for every $i$, $I_i$ is an independent set of $G$,
 \item for every $i$, $|I_i|\ge k$, and 
 \item for every $i$, $I_{i+1}$ can be obtained from $I_i$ using at most one \TAR-step.
\end{itemize}
Observe that $I\tar_k^G J$ if and only if there exists a $k$-\TARseq\ in $G$ from $I$ to $J$.
Note that we allow that $I_i=I_{i+1}$, in order to avoid discussing trivial cases in our proofs.

Our results also apply to the {\em token jumping (TJ)} model: for a graph $G$ and integer $k$, the vertex set of the graph $\TJ_k(G)$ is the set of all independent sets of size {\em exactly} $k$ in $G$. Two distinct independent sets $I$ and $J$ are adjacent in $\TJ_k(G)$ if there exist vertices $u\in I$ and $v\in J$ such that $I\bs \{u\}=J\bs \{v\}$. We say that $J$ is obtained from $I$ by {\em jumping a token from $u$ to $v$.} 
Analogously to before, this defines \TJseq s from $I$ to $J$, and we write $I\tj^G J$ if a \TJseq\ from $I$ to $J$ exists. 
Kami\'{n}ski et al showed that the TAR-model generalizes the TJ-model, in the following way: 
\begin{lem}[\cite{KMM12}]
\label{lem:TJisTAR}
Let $A$ and $B$ be two \indset s of a graph $G$, with $|A|=|B|=\ell$. 
Then for any $k\in \mathbb{N}$, there exists an $(\ell-1)$-\TARseq\ from $A$ to $B$ of length at most $2k$ if and only if there exists a \TJseq\ from $A$ to $B$ of length at most $k$.
\end{lem}
We remark that the TAR-model as defined in~\cite{KMM12} is a little more restricted: for our algorithms, it is essential to consider the case where the token lower bound $k$ is equal to the size of the initial \indset s $A$ and $B$, whereas in~\cite{KMM12}, only the case where $k<\min \{|A|,|B|\}$ is considered.

\paragraph{Cographs and cotree decompositions}

For an illustration of the following definitions, see Figure~\ref{fig:cotree}.
A {\em generalized cotree} is a binary tree $T$ with root $r$, together with 
\begin{itemize}
 \item a partition of the nonleaf vertices into {\em union nodes} and {\em join nodes}, and 
 \item 
 a graph $G_u$ for every leaf $u$ of $T$, such that for any two leaves $u$ and $v$, the graphs $G_u$ and $G_v$ are vertex and edge disjoint.
\end{itemize}
Vertices of $T$ are called {\em nodes}. 
For every nonleaf node $u$, the two children are ordered; they are called the {\em left child} and {\em right child} of $u$.
With every node $u\in V(T)$ we associate a graph $G_u$ in the following way:
for leaves $u$, $G_u$ is as given.
Otherwise, $u$ has two child nodes; denote these by $v$ and $w$.
If $u$ is a union node, then $G_u$ is the {\em disjoint union} of $G_v$ and $G_w$. 
If $u$ is a join node, then $G_u$ is obtained by taking the {\em complete join} of $G_v$ and $G_w$. This operation is defined as follows: start with the disjoint union of $G_v$ and $G_w$, and add edges $yz$ for every combination of $y\in V(G_v)$ and $z\in V(G_w)$.
For a node $u\in V(T)$, we denote $V_u=V(G_u)$.
A generalized cotree $T$ is called a {\em cotree} if for every leaf $v\in V(T)$, the graph $G_v$ consists of a single vertex. Such a leaf is called a {\em trivial leaf}.

Let $T$ be a (generalized) cotree, with root $r$.
For a graph $G$, we say that $T$ is a {\em (generalized) cotree for $G$} if $G_r=G$. 
A graph $G$ is called a {\em cograph} if there exists a cotree for $G$.
Let $\GG$ be a graph class. We say that a generalized cotree $T$ for a graph $G$ is a {\em cotree decomposition of $G$ into $\GG$-graphs} if for every leaf $v\in V(T)$, the graph $G_v\in \GG$. For instance, we will consider cotree decompositions into 
chordal graphs.

\section{Example and Proof Outline}
\label{sec:outline}

In this section, we will give an example, and use it to introduce the techniques and notions that will be used in the proofs. We will end with an outline of the algorithm, and overview of the paper.

\paragraph{Example}

\begin{figure}
\centering
\scalebox{1.2}{$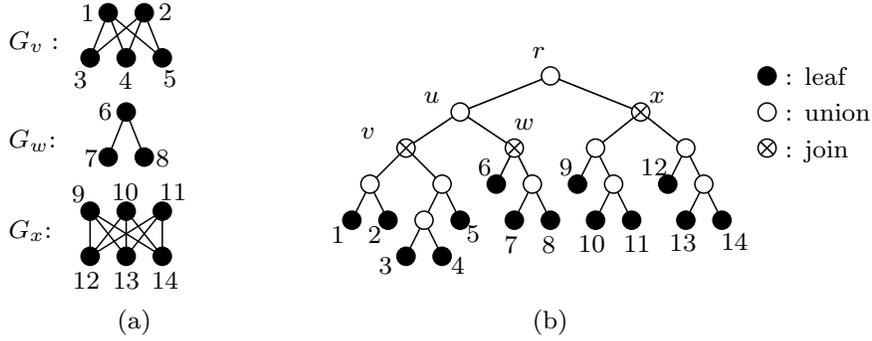$}
\caption{(a) A cograph $G$ with components $G_v$, $G_w$ and $G_x$, and (b) a cotree $T$ of $G$. Leaves of $T$ are labeled with the corresponding vertex number of $G$.}
\label{fig:cotree}
\end{figure}
In Figure~\ref{fig:cotree}, a cograph $G$ together with a cotree $T$ of $G$ is shown. The root of $T$ is $r$, and $V(G)=\{1,\ldots,14\}$. 
The graph $G$ has three components, which are $G_v$, $G_w$ and $G_x$. 

\begin{figure}
\centering
\scalebox{1.2}{$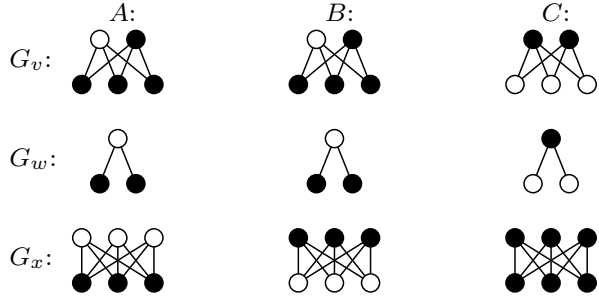$}
\caption{A cograph $G$, with independent sets $A$, $B$ and $C$ indicated by the white vertices. Any $5$-\TARseq\ from $A$ to $B$ must visit $C$, use all vertices of $G$, and has length at least 24.}
\label{fig:longTARseq}
\end{figure}
In Figure~\ref{fig:longTARseq}, three independent sets $A$, $B$ and $C$ are shown for the cograph $G$ from Figure~\ref{fig:cotree}. 
In order to go from $A$ to $B$ in $\TAR_5(G)$, an \indset\ must be visited which has no tokens on the component $G_x$, and therefore at least five tokens on the other two components. The only such \indset\ of $G$ is $C$. Using similar observations, it can be seen that there the {\em shortest} $5$-\TARseq\ from $A$ (or $B$) to $C$ is unique up to symmetries, and has length twelve (six additions and six deletions). Hence the shortest $5$-\TARseq\ from $A$ to $B$ has length 24.

\paragraph{Proof Outline and Definitions}
For two \indset s $A$ and $B$ of a graph $G$, both with size at least $k$, we will characterize whether $A\tar_k^G B$, using a (generalized) cotree for $G$. This requires the following notion.

\begin{defi}
\label{def:freedom}
Let $T$ be a generalized cotree for a graph $G$, $I$ be an independent set of $G$, and $k\le |I|$.
For $v\in V(T)$, define $\freedom^I_k(v)=\min |J\cap V_v|$ over all \indset s $J$ of $G$ with $I\tar_k^G J$.
\end{defi}

For instance, in the example from Figure~\ref{fig:longTARseq}, $\freedom^A_5(x)=0=\freedom^B_5(x)$, and this fact is essential for concluding that $A\tar_5^G B$ in this case. In general, the following theorem characterizes whether $B$ is reachable from $A$, using the values from Definition~\ref{def:freedom}.

\newcounter{THM:main_combin}
\setcounter{THM:main_combin}{\value{thm}}
\begin{thm}
\label{thm:NEW_main_combin}
Let $T$ be a generalized cotree for a graph $G$. 
Let $A$ and $B$ be two \indset s of $G$ of size at least $k$.
Then $A\tar_k^G B$ if and only if
\begin{enumerate} 
 \item
 \label{it:one}
 for all nodes $u\in V(T)$, $\freedom^A_k(u)=\freedom^B_k(u)$, and
 \item
 \label{it:two}
 for all leaves $u\in V(T)$, $(A\cap V_u) \tar_{\ell}^{G_u} (B\cap V_u)$, where $\ell=\freedom^A_k(u)$.
\end{enumerate}
\end{thm}

The forward direction of the statement is straightforward: if $A\tar_k^G B$, then since $\tar_k^G$ is an equivalence relation, any \indset\ $J$ is reachable from $A$ if and only if it is reachable from $B$. It follows that $\freedom^A_k(v)=\freedom^B_k(v)$ for all $v\in V(T)$. The second property follows by restricting all \indset s in a $k$-\TARseq\ from $A$ to $B$ to the subgraph $G_v$ for any leaf $v\in V(T)$. By definition, these all have size at least $\ell=\freedom_k^A(v)$, so this yields an $\ell$-\TARseq\ from $A\cap V_v$ to $B\cap V_v$ for $G_v$. For more details, see Section~\ref{sec:summary}, where Theorem~\ref{thm:NEW_main_combin} is proved.

In order to efficiently decide whether $A\tar_k^G B$, it remains to compute the values $\freedom_k^I(v)$ for all $v\in V(T)$ and $I=A,B$. How this can be done is shown in Section~\ref{sec:topdown}.

In the example from Figure~\ref{fig:longTARseq}, it holds that $\freedom_5^A(x)=0$. This is because on the subgraph $G_u$, which is the disjoint union of components $G_v$ and $G_w$, it is possible to reconfigure from the initial \indset\ $A$ to an \indset\ with at least five tokens on $G_u$, while keeping at least two tokens on $G_u$ throughout. This indicates that in order to compute the values $\freedom_k^I(v)$, the following values must be computed, for different values of $\ell\in \{0,\ldots,k\}$.

\begin{defi}
\label{def:RIS}
Let $T$ be a cotree for $G$, and $I$ be an independent set of $G$.
For $v\in V(T)$ 
and $\ell\in \{0,\ldots,|I\cap V_v|\}$, denote by $\RIS^I_\ell(v)$ the maximum of $|J|$ over all independent sets $J$ of $G_v$ with $(I\cap V_v)\tar_{\ell}^{G_v} J$.
\end{defi}

Note that the value $\RIS^I_\ell(v)$ depends only on the situation in the subgraph $G_v$; not on the entire graph. This is in contrast to the values $\freedom^I_k(v)$. 
Observe also that $\RIS^I_0(v)=\alpha(G_v)$ (regardless of the choice of $I$).

It is not obvious how to compute the values $\RIS^I_{\ell}(u)$. For the example from Figure~\ref{fig:longTARseq}, concluding that $\RIS^A_2(u)=5$ requires studying the following 2-\TARseq\ for $G_u$. 
We start with one token on both $G_v$ and $G_w$. One token can be added on $G_v$. This allows removing the token from $G_w$, and subsequently moving to a better configuration, with two tokens on $G_w$. This in turn allows removing all tokens from $G_v$, and subsequently moving to a better configuration, with three tokens on $G_v$.
A sequence of this type is called a {\em cascading sequence}. 
Informally, in such a sequence, we have a join node $u$ with children $v$ and $w$, and alternatingly move between on one hand a large \indset\ on $v$ and a small \indset\ on $w$ and on the other hand a large \indset\ on $w$ and small \indset\ on $v$. The goal is to obtain ever larger \indset s until no more improvements can be made.
In Section~\ref{ssec:RISrules}, we will show how to compute the values $\RIS^I_{\ell}(v)$. This is done by characterizing the outcome of such cascading sequences, using {\em maximum $\ell$-stable tuples}. 

The values $\RIS^I_{\ell}(u)$ for a node $u$ with children $v$ and $w$ can be computed using only the values $\RIS^I_{\ell'}(v)$ and $\RIS^I_{\ell'}(w)$ for different choices of $\ell'$. Hence these values can be computed using a {\em bottom up} dynamic programming algorithm, which starts at the leaves of the cotree.
Next, the rules from Section~\ref{sec:topdown} for computing the values $\freedom^I_k(u)$ can be used. As indicated by their definitions, computing these values requires considering the entire graph. Therefore this must be done using a {\em top down} dynamic programming algorithm, which starts at the root node of $T$.
Together with Theorem~\ref{thm:NEW_main_combin}, this yields our algorithm for deciding whether $A\tar_k^G B$.
Our main algorithmic result is summarized in the next theorem, which is proved in Section~\ref{sec:summary}.

\newcounter{THM:main_alg}
\setcounter{THM:main_alg}{\value{thm}}
\begin{thm}
\label{thm:main_alg}
Let $T$ be a generalized cotree for a graph $G$ on $n$ vertices, let $k\in \mathbb{N}$ and let $A$ and $B$ be \indset s of $G$. 
If for every nontrivial leaf $v\in V(T)$ and relevant integer $\ell$, 
\begin{itemize}
\item 
the values $\RIS^A_\ell(v)$ and $\RIS^B_\ell(v)$ are known, and 
\item 
it is known whether $(A\cap V_v)\tar_\ell^{G_v} (B\cap V_v)$, 
\end{itemize}
then in time $O(n^2)$ it can be decided whether $A\tar_k^G B$.
\end{thm}
In particular, Theorem~\ref{thm:main_alg} implies that for any two \indset s $A$ and $B$ for a {\em cograph} $G$, it can be decided in time $O(n^2)$ whether $A\tar_k^G B$.

In Section~\ref{sec:diameter}, we will give an upper bound for the length of a shortest $k$-\TARseq\ between two \indset s $A$ and $B$. The above example shows that to go from $A$ to $B$, it may be necessary to put tokens on vertices that are neither in $A$ nor in $B$. Nevertheless, we can show that for a commonly reachable \indset\ $C$, there exists a $k$-\TARseq\ from $A$ (resp.\ $B$) to $C$ that for every vertex $v\in V(G)$, adds a token on $v$ at most once. This shows that there exists a $k$-\TARseq\ from $A$ to $B$ of length at most $4n-|A|-|B|$. 

For all of our proofs, an essential fact is that for every node $u$, the vertex set $V_u$ is a module of $G$. We will first give lemmas related to \indset\ reconfiguration and modules in Section~\ref{sec:ModuleLemmas}.

\section{Module Lemmas}
\label{sec:ModuleLemmas}

A {\em module} of a graph $G$ is a set $M\subseteq V(G)$ such that for every $v\in V(G)\bs M$, either $M\subseteq N(v)$ or $M\cap N(v)=\emptyset$. In other words: for every pair $u,v\in M$, $N(u)\bs M=N(v)\bs M$. Note that we will also consider $V(G)$ to be a (trivial) module of $G$.
We will often use the following simple property of cographs.
\begin{propo}
\label{propo:module}
Let $T$ be a cotree of $G$. Then for any $v\in V(T)$, $V_v$ is a module of $G$.
\end{propo}

Modules are very useful for independent set reconfiguration, since to some extent, we can reconfigure within the module and outside of the module independently. 
The following two lemmas make this more precise, and present two useful properties for the proofs below.
These two lemma proofs also introduce proof techniques related to \TARseq s that will be used often below. Later, we will however not apply them in the same level of detail again.

\begin{lem}
\label{lem:moduleA}
Let $M$ be a module of a graph $G$, let $k$ and $y$ be integers, and let $A$ be an \indset\ of $G$, with $|A\cap M|\ge \max\{1,y\}$ and $|A|\ge k$. Denote $H=G[M]$.
If there exists an \indset\ $B$ of $G$ with $A\tar^G_k B$ and $|B\cap M|\le y$, and if there exists an \indset\ $C$ of $H$ with $(A\cap M)\tar^H_y C$, then there exists an \indset\ $D$ of $G$ with $A\tar^G_k D$ and $D\cap M=C$.
\end{lem}

\PF
Denote $A_M=A\cap M$, and $A_{\Mc}=A\bs M$.
First consider the case that $|A_M|=y$. 
Informally, we can then simply apply the same vertex additions and removals from the $y$-\TARseq\ from $A_M$ to $C$ to the entire \indset\ $A$, and this way maintain an independent set throughout.

Formally, let $I_0,\ldots,I_p$ be an $y$-\TARseq\ for $H$ from $A_M$ to $C$. Define $I'_i=I_i\cup A_{\Mc}$ for all $i$. Then $I'_0,\ldots,I'_p$ is the desired $k$-\TARseq\ from $A$ to an \indset\ $D$ of $G$ with $D\cap M=C$. Indeed,
\begin{itemize}
 \item for every $i$, both $I_i$ and $A_{\Mc}$ are \indset s. 
 Since $M$ is a module and $|A_M|\ge 1$, $A_{\Mc}$ contains no vertices that are adjacent to any vertex in $M$, so $I'_i$ is again an independent set of $G$.
 \item Since $|A_M|=y$, we have $|A_{\Mc}|\ge k-y$. By definition, for every $i$ it holds that $|I_i|\ge y$, and thus $|I'_i|\ge y+k-y\ge k$.
 \item Clearly, every $I'_{i+1}$ can be obtained from $I'_i$ using at most one \TAR-step.
\end{itemize}

\medskip 
In the remaining case, we may assume that $|A_M|\ge y+1$.
Consider a {\em shortest} $k$-\TAR-seq\ $S=J_0,\ldots,J_q$ from $A$ to {\em any} \indset\ $B$ of $G$ with $|B\cap M|\le y$.
So for every $i$ with $i<q$, $|J_i\cap M|\ge y+1$, 
and $B=J_q$ is obtained from $J_{q-1}$ by removing a vertex from $M$.
Since $M$ is a module and $J_{q-1}$ is an \indset, this implies that no vertex in $B\bs M$ is adjacent to any vertex in $M$. 
Denote $B_{\Mc}=B\bs M$. 

Informally, we can now reverse the \TARseq\ $S$, but ignore every token addition or removal  on $V(G)\bs M$. This yields a $k$-\TARseq\ for $G$, from $B$ to $A_M\cup B_{\Mc}$. Since $|B_{\Mc}|\ge k-y$, we can now apply the token additions and removals from the \TARseq\ from $A_M$ to $C$ to this \indset, similar to above, and obtain the desired \indset\ $D=C\cup B_{\Mc}$. Combining these three $k$-\TARseq s shows that there is a $k$-\TARseq\ from $A$ to $D$. We now define this more precisely, and verify that these are indeed \TARseq s.

For every $i$, denote $J'_i=(J_i\cap M)\cup B_{\Mc}$. Consider the sequence $S'=J'_q,\ldots,J'_0$. The argue that this is a $k$-\TARseq\ from $B$ to $A_M\cup B_{\Mc}$:
\begin{itemize}
 \item As observed above, no vertex in $B_{\Mc}$ is adjacent to any vertex in $M$. Hence for every $i$, $J'_i$ is an independent set.
 \item Recall that for every $i$, $|J_i\cap M|\ge y$, and $|B_{\Mc}|\ge k-y$, so $|J'_i|\ge k$.
 \item Clearly, consecutive sets in the sequence can be obtained from each other by at most one \TAR-step.
\end{itemize}
Analog to the first part of the proof, one can show that there exists a $k$-\TARseq\ $S''$ for $G$ from $A_M\cup B_{\Mc}$ to $C\cup B_{\Mc}$. Combining the sequences $S$, $S'$ and $S''$ shows that $A\tar^G_k (C\cup B_{\Mc})$, which proves the statement.
\QED

Using similar techniques, we can prove the next lemma. 
This applies to the case where one module $M$ can be partitioned into two sets $M_1$ and $M_2$, with no edges between them, which therefore are also modules.

\begin{lem}
\label{lem:moduleB} 
Let $M$ be a module of a graph $G$, such that $M$ can be partitioned into two sets $M_1$ and $M_2$ with no edges between $M_1$ and $M_2$.
Let $A$ be an \indset\ of $G$, let $B_1$ be an \indset\ of $G$ with $A\tar^G_k B_1$, that maximizes $|B_1\cap M_1|$ among all such sets, and let $B_2$ be an \indset\ of $G$ with $A\tar^G_k B_2$.

Then there exists an \indset\ $C$ of $G$ with $A\tar^G_k C$ and $C\cap M_i=B_i\cap M_i$, for $i\in \{1,2\}$.
\end{lem}

\PF
If $B_1\cap M_1=\emptyset$, then by definition of $B_1$, it also holds that $B_2\cap M_1=\emptyset$, and therefore chosing $C=B_2$ proves the statement.
So now suppose that $B_1\cap M_1\not=\emptyset$.

Since the relation $\tar^G_k$ is an equivalence relation, we conclude that there exists a $k$-\TARseq\ $S=I_0,\ldots,I_p$ from $B_1$ to $B_2$. We will use $S$ to show that there exists an \indset\ $C$ of $G$ with $B_1\tar^G_k C$ and $C\cap M_i=B_i\cap M_i$, for $i\in \{1,2\}$. Combining this with $A\tar^G_k B_1$ shows that also $A\tar^G_k C$, which proves the statement. 

First suppose that $S$ contains an \indset\ that contains no vertices of $M$.
Then let $I_i$ be the first such independent set, so $i\ge 1$ and $I_i$ is obtained from $I_{i-1}$ by removing a token from $M$. Since $M$ is a module and $I_{i-1}$ is an \indset, $I_i$ therefore contains no vertex that is adjacent to any vertex in $M$. We can then simply add the vertices in $B_1\cap M_1$ and $B_2\cap M_2$ to $I_i$ in any order, to obtain the desired \indset\ $C$ (recall that there are no edges between $M_1$ and $M_2$). 
Combining the \TARseq s from $A$ to $B_1$, from $B_1$ to $I_i$, and from $I_i$ to $C$ shows that $A\tar^G_k C$.

So now we may assume that every \indset\ $I_i$ in the sequence $S$ contains at least one vertex of $M$.
Then we modify $S$ as follows: we ignore all token additions and removals on $M_1$.
We argue that this is a $k$-\TARseq\ from $B_1$ to $C=(B_1\cap M_1)\cup (B_2\bs M_1)$.
More precisely, for every $i$ define $I'_i=(B_1\cap M_1)\cup (I_i\bs M_1)$, and consider the sequence $S'=I'_0,\ldots,I'_p$. We argue that $S'$ is a $k$-\TARseq\ for $G$:
\begin{itemize}
\item 
Suppose to the contrary that there exists an $i$ such that $I'_i$ is not an \indset. 
Let $i$ be the minimum index with this property.
Then $I_i$ is obtained from $I_{i-1}$ by adding a vertex $y$ that is adjacent to some vertex in $B_1$. Because vertices in $M_2$ are not adjacent to vertices in $M_1$, it follows that $y\not\in M$. 
Since $M$ is a module, $y$ is adjacent to every vertex in $M$. But $I_i$ contains at least one vertex of $M$, contradicting that it is an \indset. We conclude that every $I'_i$ is an \indset. 
\item
Let $y=|B_1\cap M_1|$. Since every \indset\ $I_i$ is also reachable from $A$, from the definition of $B_1$ it follows that $|I_i\cap M_1|\le y$. Therefore, $|I_i\bs M_1|\ge k-y$. It follows that for every $i$, $|I'_i|\ge y+k-y=k$.
\item
Clearly, consecutive sets in the sequence $S'$ can be obtained from each other using at most one \TAR-step.
\end{itemize}
We conclude that $B_1\tar^G_k C$. Combined with $A\tar^G_k B_1$, it follows that $A\tar^G_k C$.\QED

\section{Dynamic Programming Rules}
\label{sec:DP}

\subsection{Cascading Sequences}
\label{ssec:casc}

\begin{figure}
\centering
\scalebox{1.2}{$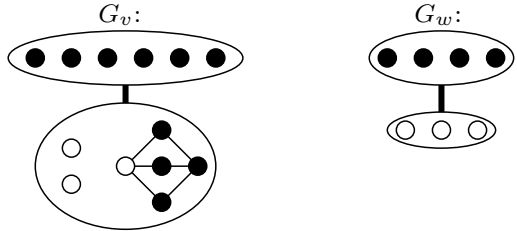$}
\caption{A cograph $G_u$ with components $G_v$ and $G_w$, and \indset\ $I$ consisting of the white vertices.}
\label{fig:cascading}
\end{figure}

In Section~\ref{ssec:RISrules} below we will give dynamic programming rules for computing the values $\RIS_{\ell}^I(u)$ for all nodes $u\in V(T)$. 
Recall that $\RIS^I_\ell(u)=\max |J|$ where the maximum is taken over all independent sets $J$ of $G_u$ with $(I\cap V_u)\tar_{\ell}^{G_u} J$.
For trivial leaves and join nodes, the rules are straightforward. As discussed in Section~\ref{sec:outline}, the computation for union nodes is more complicated, and requires studying the outcome of certain $\ell$-\TARseq s in $G_u$, which we will informally call {\em cascading sequences}.

We first introduce these informally, using the example shown in Figure~\ref{fig:cascading}. This figure depicts a cograph $G_u$ which is obtained by taking the disjoint union of two cographs $G_v$ and $G_w$. In this figure, a bold line between two encircled sets $V_1$ and $V_2$ of vertices means that edges $xy$ are present between every $x\in V_1$ and $y\in V_2$. This corresponds to a complete join of $G[V_1]$ and $G[V_2]$. 
Let $I$ be the \indset\ of $G_u$ consisting of the white vertices.
In Table~\ref{tab:RISexample}, the values $\RIS^I_{\ell}(v)$ and $\RIS_{\ell}^I(w)$ are given for every $\ell\in \{0,\ldots,3\}$. 
(These values can easily be verified. See also Figure~\ref{fig:SUseqs} for examples of maximum \indset s that are reachable from $I\cap V_x$ in $\TAR_\ell(G_x)$ for various values of $\ell$, and $x\in \{v,w\}$.)

Let the {\em type} of an independent set $J$ of $G_u$ be $(|J\cap V_v|,|J\cap V_w|)$.
If it is required to keep at least $\ell=6$ tokens on $G_u$ throughout, then from the initial \indset\ $I$, which is of type $(3,3)$, we can go to an \indset\ of type $(\RIS_3^I(v),\RIS_3^I(w))=(4,3)$. 
This holds by definition of $\RIS_3^I(v)$ and $\RIS_3^I(w)$, and because we can reconfigure in both components independently, as long as at least three tokens remain on both sides.
From this, we could go to a \indset\ of type $(4,2)$, but this does not enable further improvements. So we conclude that $\RIS_6^I(u)=4+3=7$.
If $\ell=5$, then observe that we can go from the initial \indset\ of type $(3,3)$ to one of type $(2,3)$, and subsequently to one of type $(\RIS_2^I(v),3)=(5,3)$. 
Next, we can visit \indset s of types $(5,0)$ and $(5,\RIS_0^I(w))=(5,4)$. We could then go to one of type $(1,4)$, but since $\RIS_1^I(v)=5$, this yields no improvement.
We conclude that $\RIS_5^I(u)=5+4=9$.
Finally, if $\ell\le 4$, then we can visit \indset s of types $(3,3)$, $(4,3)$, $(4,0)$, $(4,4)$, $(0,4)$, $(6,4)$, in this order, using similar arguments. Since $6+4=10=\alpha(G_u)$, no further improvements are possible, so $\RIS_\ell^I(u)=10$ for all $\ell\le 4$.
This yields the values $\RIS^I_\ell(u)$ shown in Table~\ref{tab:RISexample}.
Note that we can deduce these values using only the previous two columns of the table; without considering other details about the graph.

\begin{table}
\[
\begin{array}{|l|lll|}

 \ell: 	& \RIS^I_{\ell}(v):	& \RIS^I_{\ell}(w):	& \RIS^I_{\ell}(u):\\
 \hline
 0	& 6			& 4			& 10\\
 1	& 5			& 3			& 10\\
 2	& 5			& 3			& 10\\
 3	& 4			& 3			& 10\\
 4	&			&			& 10\\
 5	&			&			& 9\\
 6	&			&			& 7\\
 
\end{array}
\]
\caption{The values $\RIS^I_\ell(x)$ for $x\in \{u,v,w\}$ and $\ell\in \{0,\ldots,|I\cap V_x|\}$.}
\label{tab:RISexample}
\end{table}

Below we will prove that the values computed this way are correct. However, we will not formalize cascading sequences, but instead characterize their outcome. We will define {\em maximum $\ell$-stable tuples $(x,y)$} for each $\ell\in \{0,\ldots,|I\cap V_u|\}$, and show that $x=\min |J\cap V_v|$ and $y=\min |J\cap V_w|$, where in both cases the minimum is taken over all \indset s $J$ of $G_u$ with 
$(I\cap V_u)\tar_{\ell}^{G_u} J$. So for the example above and $\ell=6,5,4$, these tuples can be verified to be $(3,2)$, $(1,0)$ and $(0,0)$ respectively. 
(As indicated by the above cascading sequences.)
Next, we will show that $\RIS_\ell^I(u)=\RIS_x^I(v)+\RIS_y^I(w)$, where $(x,y)$ is the maximum $\ell$-stable tuple. The maximum $\ell$-stable tuple can easily be computed from its definition, given below.

\subsection{Bottom Up Dynamic Programming Rules}
\label{ssec:RISrules}

Throughout this section, $T$ denotes a generalized cotree of $G$ and $I$ denotes an \indset\ of $G$.
The following property follows easily from the definition of $\RIS_{\ell}^I(u)$, and will often be used in this section.
\begin{propo}
\label{propo:RISmonotone}
Let $u\in V(T)$. For any two integers $x,y$ with $0\le x\le y\le |I\cap V_u|$, it holds that $\RIS_x^I(u)\ge \RIS_y^I(u)$.
\end{propo}

For trivial leaf nodes, the computation of these values is easy:

\begin{propo}
\label{propo:RISleaf}
Let $u\in V(T)$ be a trivial leaf node. Then $\RIS^I_\ell(u)=1$ for all $\ell$.
\end{propo}

For join nodes, the computation of $\RIS_{\ell}^I(u)$ is still relatively straightforward. Note that for any join node $u$ and \indset\ $I$, $u$ has a child $w$ with $V_w\cap I=\emptyset$.

\begin{propo}
\label{propo:RISjoin}
Let $u\in V(T)$ be a join node. Let $w$ be a child of $u$ with $I\cap V(G_v)=\emptyset$, and let $v$ be the other child of $u$.
\begin{itemize}
 \item $\RIS^I_\ell(u)=\RIS^I_\ell(v)$ for all $\ell\ge 1$, and
 \item $\RIS^I_0(u)=\max\{\RIS^I_0(v),\RIS^I_0(w)\}$.
\end{itemize}
\end{propo}
\PF
Because all edges are present between $G_v$ and $G_w$, a maximum independent set of $G_u$ is either a maximum \indset\ of $G_v$ or of $G_w$, so $\RIS^I_0(u)=\alpha(G_u)=\max\{\alpha(G_v),\alpha(G_w)\}=\max\{\RIS^I_0(v),\RIS^I_0(w)\}$. 
Now consider the case $\ell\ge 1$, and thus $|I\cap V_u|\ge 1$. 
Then initially all tokens of $I$ are on the child $G_v$. 
As long as there is at least one token on $G_v$, no tokens can be added to $G_w$. 
So essentially, $G_w$ can be ignored, and thus $\RIS^I_\ell(u)=\RIS^I_\ell(v)$.
\QED

For union nodes $u$, 
we will show that the value of $\RIS_{\ell}^I(u)$ can be characterized using {\em maximum stable tuples}, which are defined as follows. 

\begin{defi}
For a union node $u$ with left child $v$ and right child $w$, independent set $I\subseteq V(G)$ and integer $\ell\le |I\cap V_u|$, call a tuple $(x,y)$ of integers with $x\le |I\cap V_v|$ and $y\le |I\cap V_w|$ {\em $\ell$-stable} if 
\[ 
x=\max\{0,\ell-\RIS^I_y(w)\} \mbox{ and } y=\max\{0,\ell-\RIS^I_x(v)\}.
\]
Call an $\ell$-stable tuple $(x,y)$ {\em maximum} if there is no $\ell$-stable tuple $(x',y')$ with $x'\ge x$, $y'\ge y$ and $(x,y)\not=(x',y')$.
\end{defi}

In the remainder of this section, we will first prove that for every $\ell$, there exists a unique maximum $\ell$-stable tuple $(x,y)$, and characterize this tuple (Lemma~\ref{lem:UniqueStableCharacterization} below). Using this characterization, we can show that for a union node $u$ with children $v$ and $w$ and any $\ell$, $\RIS^I_{\ell}(u)=\RIS^I_x(v)+\RIS^I_y(w)$, where $(x,y)$ is the unique maximum $\ell$-stable tuple (Lemma~\ref{lem:RISunion} below).

\begin{lem}
\label{lem:UniqueStableCharacterization}
Let $u\in V(T)$ be a union node, with left child $v$ and right child $w$.  
For $\ell\in \{0,\ldots,|I\cap V_u|\}$, let 
$x=\min |J\cap V_v|$ and $y=\min |J\cap V_w|$, where in both cases the minimum is taken over all \indset s $J$ of $G_u$ with $(I\cap V_u) \tar_{\ell}^{G_u} J$.
Then $(x,y)$ is the unique maximum $\ell$-stable tuple for $I$ and $u$.
\end{lem}

Before we can prove Lemma~\ref{lem:UniqueStableCharacterization}, we first need to prove a number of other statements. 
These statements will refer to notations $I$, $u$, $v$, $w$, $x$, $y$, $\ell$ as defined in Lemma~\ref{lem:UniqueStableCharacterization}.
In addition, we will denote $I_u=I\cap V_u$.

\begin{propo}
\label{propo:ClaimA}
Consider a \TARseq\ $S=J_0,\ldots,J_p$ in $\TAR_{\ell}(G_u)$ with $J_0=I_u$.
Let $x^*=\min_i |J_i\cap V_v|$ and $y^*=\min_i |J_i\cap V_w|$.
Then $\RIS_{x^*}^I(v)\ge |J_p\cap V_v|$ and $\RIS_{y^*}^I(w)\ge |J_p\cap V_w|$.
\end{propo}

\PF
Consider the sequence $S'=J'_0,\ldots,J'_p$ with $J'_i=J_i\cap V_w$ for all $i$. 
This is a $y^*$-\TARseq\ for $G_w$ that ends with $J'_p$. 
So by definition, $\RIS_{y^*}^I(w)\ge |J'_p|$.
The proof of the other statement is analog.
\QED

Since $x$ and $y$ (as defined in Lemma~\ref{lem:UniqueStableCharacterization}) provide lower bounds for $x^*$ and $y^*$ respectively (as defined in Proposition~\ref{propo:ClaimA}), we conclude:

\begin{corol}
\label{cor:RISisUB}
For every $J$ with $I_u\tar_{\ell}^{G_u} J$, it holds that $\RIS_x^I(v)\ge |J\cap V_v|$ and $\RIS_y^I(w)\ge |J\cap V_w|$.
\end{corol}

Using Proposition~\ref{propo:ClaimA}, we can draw the following two conclusions. 

\begin{propo}
\label{propo:eqone}
$x\ge \ell-\RIS_{y}^I(w)$ and $y\ge \ell-\RIS_{x}^I(v)$.
\end{propo}
\PF
Consider an $\ell$-\TARseq\ $J_0,\ldots,J_p$ for $G_u$ from $I_u$ to an \indset\ $J_p$ with $|J_p\cap V_v|=x$.\
Let $y^*=\min_i |J_i\cap V_w|$, so $y^*\ge y$.
By Proposition~\ref{propo:ClaimA}, 
$\RIS_{y^*}^I(w) \ge |J_p\cap V_w| \ge \ell-|J_p\cap V_v|=\ell-x$.
Using Proposition~\ref{propo:RISmonotone} and $y^*\ge y$, it follows that $\RIS_{y}^I(w)\ge \ell-x$ holds as well. 
The other inequality is proved analogously.\QED

\begin{lem}
\label{lem:eqtwo}
For any $\ell$-stable tuple $(x',y')$, it holds that
$x\ge x'$ and $y\ge y'$.
\end{lem}
\PF
Suppose to the contrary that there exists an $\ell$-\TARseq\ for $G_u$ from $I_u$ to some \indset\ $J$ with $|J\cap V_v|<x'$ or $|J\cap V_w|<y'$. 
Consider a {\em shortest} $\ell$-\TARseq\ $S=J_0,\ldots,J_p$ of this kind, and assume w.l.o.g. this ends with $J_p$ with $|J_p\cap V_v|=x'-1$. This implies that $x'\ge 1$, and therefore $x'=\ell-\RIS_{y'}^I(w)$ (since $(x',y')$ is stable).
It follows that $|J_p\cap V_w|\ge \ell-|J_p\cap V_v|=\ell-x'+1=\RIS_{y'}^I(w)+1$, so
\begin{equation}
\label{eq:X} 
|J_p\cap V_w|\ge \RIS_{y'}^I(w)+1.
\end{equation}
Combining this with the trivial lower bounds $\RIS_{y'}^I(w)\ge |I\cap V_w|\ge y'$ we obtain
\[ 
|J_p\cap V_w|\ge y'+1.
\] 
Let $y^*=\min_i |J_i\cap V_w|$, and choose $i$ accordingly such that $|J_i\cap V_w|=y^*$.
Combining Proposition~\ref{propo:ClaimA} with~(\ref{eq:X}) yields
\[
\RIS_{y^*}^I(w)\ge |J_p\cap V_w|\ge \RIS_{y'}^I(w)+1. 
\]
It follows that $y^*<y'$ (Proposition~\ref{propo:RISmonotone}). 
So $|J_p\cap V_w|\ge y'+1>y^*+1=|J_i\cap V_w|+1$, and thus $i<p$.
But then the subsequence of $S$ that ends with $J_i$ satisfies $|J_i\cap V_w|=y^*<y'$, and this is a strictly shorter sequence than $S$, a contradiction with the choice of $S$.
\QED

\begin{propo}
\label{propo:eqthree}
There exists an \indset\ $J_1$ of $G_u$ with $|J_1\cap V_v|=\RIS_{x}^I(v)$ and $I_u\tar_{\ell}^{G_u} J_1$, and there exists an \indset\ $J_2$ of $G_u$ with $|J_2\cap V_w|=\RIS_{y}^I(w)$ and $I_u\tar_{\ell}^{G_u} J_2$.
\end{propo}

\PF
We prove the second statement. The proof of the first statement is analog.
If $I_u\cap V_w=\emptyset$, then we can simply add vertices from a maximum \indset\ $C$ of $G_w$ to $I_u$, one by one. Recall that $|C|=\alpha(G_w)=\RIS_0^I(w)$.
Since $G_u$ is the disjoint union of $G_v$ and $G_w$, this yields a \TARseq\ in $G_u$, from $I_u$ to a \indset\ $J_2$ with $|J_2\cap V_w|=\RIS_0^I(w)=\RIS_{y}^I(w)$.

So we may now assume that $|I_u\cap V_w|\ge 1$, and we can apply (module) Lemma~\ref{lem:moduleA}, with $I_u$ in the role of $A$, $V_w$ in the role of the module $M$ and $G_u$ in the role of the entire graph $G$. By definition of $y$, there exists an \indset\ $B$ of $G_u$ with $I_u\tar^{G_u}_{\ell} B$ and $|B\cap V_w|=y$. By definition of $\RIS_{y}^I(w)$, there exists an \indset\ $C$ of $G_w$ with $(I_u\cap V_w)\tar^{G_w}_{y} C$ and $|C|=\RIS_{y}^I(w)$. Now Lemma~\ref{lem:moduleA} shows that there exists an \indset\ $J_2$ of $G_u$ with $A\tar^{G_u}_{\ell} J_2$ and $|J_2\cap V_w|=|C|=\RIS_{y}^I(w)$.
\QED

Now we are ready to prove Lemma~\ref{lem:UniqueStableCharacterization}.

\medskip 
\noindent 

{\em Proof of Lemma~\ref{lem:UniqueStableCharacterization}:}\/
Consider $J_2$ as in~Proposition~\ref{propo:eqthree}. 
We can remove all but $\max\{0,\ell-\RIS_{y}^I(w)\}$ tokens from $G_v$, and still have at least $\ell$ tokens in total on $G_u$. This shows that $x\le \max\{0,\ell-\RIS^I_{y}(w)\}$. Analogously, $y\le \max\{0,\ell-\RIS^I_{x}(v)\}$ follows. 
Combining these inequalities with Proposition~\ref{propo:eqone} and the obvious inequalities $x\ge 0$, $y\ge 0$ shows that 
$x=\max\{0,\ell-\RIS^I_{y}(w)\}$ and $y=\max\{0,\ell-\RIS^I_{x}(v)\}$, hence the tuple $(x,y)$ is $\ell$-stable. 
Furthermore, Lemma~\ref{lem:eqtwo} shows that $(x,y)$ is a {\em maximum} $\ell$-stable tuple, and in fact the {\em only} maximum $\ell$-stable tuple.
\QED

Lemma~\ref{lem:UniqueStableCharacterization} implies in particular that there exists a unique maximum $\ell$-stable tuple for any choice of $\ell$. From now on we will now use this fact implicitly, for instance in the following lemma statement.
Now we are ready to state and prove Lemma~\ref{lem:RISunion}, which shows how the values $\RIS^I_{\ell}(u)$ can be computed for a join node $u$.

\begin{lem}
\label{lem:RISunion}
Let $u\in V(T)$ be a union node, with left child $v$ and right child $w$.  
For $\ell\in \{0,\ldots,|I\cap V_u|\}$, let $(x,y)$ be the unique maximum $\ell$-stable tuple for $I$ and $u$. Then $\RIS^I_\ell(u)=\RIS^I_x(v)+\RIS^I_y(w)$.
\end{lem}

\PF
Lemma~\ref{lem:UniqueStableCharacterization} shows that $x=\min |J\cap V_v|$ and $y=\min |J\cap V_w|$, where in both cases the minimum is taken over all \indset s $J$ of $G_u$ with $(I\cap V_u) \tar_{\ell}^{G_u} J$, so we may apply the above statements that were proved for this choice of $x$ and $y$.

For any \indset\ $J^*$ of $G_u$ with $I_u\tar_{\ell}^{G_u} J^*$, Corollary~\ref{cor:RISisUB} shows that $|J^*|=|J^*\cap V_v|+|J^*\cap V_w|\le \RIS^I_{x}(v)+\RIS^I_{y}(w)$, so
\begin{equation}
 \label{eq:eight}
 \RIS^I_{\ell}(u)\le \RIS^I_{x}(v)+\RIS^I_{y}(w).
\end{equation}
Now it suffices to prove that
\begin{equation}
\label{eq:seven}
\RIS^I_{\ell}(u)\ge \RIS^I_{x}(v)+\RIS^I_{y}(w).  
\end{equation}
To this end, we will show that (module) Lemma~\ref{lem:moduleB} can be applied, with $G_u$ in the role of the entire graph $G$, $V_u$ in the role of the module $M$, and $V_v$ and $V_w$ in the roles of the modules $M_1$ and $M_2$, respectively (recall that $\{M_1,M_2\}$ should be a partition of $M$ with no edges between $M_1$ and $M_2$). 
We choose $I_u$ in the role of $A$. 
By Proposition~\ref{propo:eqthree}, there exists an \indset\ $J_1$ of $G_u$ with $|J_1\cap V_v|=\RIS^I_x(v)$ and $I_u\tar_{\ell}^{G_u} J_1$. 
Corollary~\ref{cor:RISisUB} shows that $J_1$ maximizes the number of vertices on $V_v$ among all reachable sets. 
Analogously, these two propositions show that there exists an \indset\ $J_2$ of $G_u$ with $|J_2\cap V_v|=\RIS^I_{y}(w)$ and $I_u\tar_{\ell}^{G_u} J_2$, that maximizes the number of vertices on $V_w$ among all reachable \indset s. 
When using $J_1$ and $J_2$ in the roles of $B_1$ and $B_2$, Lemma~\ref{lem:moduleB} shows that there exists an \indset\ $C$ of $G$ with $I_u\tar_{\ell}^{G_u} C$, $C\cap V_v=J_1\cap V_v$ and $C\cap V_w=J_2\cap V_w$. Inequality~(\ref{eq:seven}) follows since $|J_1\cap V_v|=\RIS^I_{x}(v)$ and $|J_2\cap V_w|=\RIS^I_{y}(w)$.
\QED

\subsection{Computing the Values Efficiently}
\label{ssec:fastRIScomput}

Let $u$ be a union node with children $v$ and $w$ such that for every relevant integer $\ell$, the values $\RIS^I_{\ell}(v)$ and $\RIS^I_{\ell}(w)$ are known. Then Lemma~\ref{lem:RISunion} shows that for every relevant value $\ell$, the value $\RIS^I_{\ell}(u)$ can be computed in polynomial time: try all relevant combinations $(x,y)$, verify whether they are stable, and subsequently identify the unique maximum stable tuple. However, this is not very efficient. In this section we present a more efficient method for computing the values $\RIS_{\ell}^I(u)$ for union nodes $u$.
The method is shown in Algorithm~\ref{alg:fastRIScomput}.

\begin{algorithm}
\caption{Efficiently computing all values $\RIS^I_\ell(u)$ for a union node $u$.}
\label{alg:fastRIScomput}
{\sc Input:} For a union node $u$ with children $v$ and $w$: 
 The values $|I\cap V_z|$ and $\RIS_\ell^I(z)$ for $z\in \{v,w\}$ and $\ell\in \{0,\ldots,|I\cap V_z|\}$.\\
{\sc Output:} Values $\RIS_\ell^I(u)$ for all $\ell\in \{0,\ldots,|I\cap V_u|\}$.
\begin{ntabbing}
\qquad \= \qquad \= \kill
$a:=|I\cap V_v|$\label{la:1}\\
$b:=|I\cap V_w|$\label{la:2}\\
$\ell:=a+b$\label{la:3}\\
{\bf while} $\ell\ge 0$ {\bf do}\label{la:4}\\
\> {\bf repeat}\label{la:5}\\
\> \> $a':=a$\label{la:6}\\
\> \> $b':=b$\label{la:7}\\
\> \> $a:=\max\{0,\ell-\RIS_{b'}^I(w)\}$\label{la:assign1}\\
\> \> $b:=\max\{0,\ell-\RIS_{a'}^I(v)\}$\label{la:assign2}\\
\> {\bf until} $a=a'$ and $b=b'$\label{la:10}\\
\> $\RIS_{\ell}^I(u):=\RIS_a^I(v)+\RIS_b^I(w)$\label{la:assignRIS}\\
\> $\ell:=\ell-1$\label{la:decell}\\
{\bf endwhile}\label{la:13}
\end{ntabbing}
\end{algorithm}

To prove that Algorithm~\ref{alg:fastRIScomput} is correct, we need the following invariant.

\begin{propo}
\label{propo:abovestable}
At any time during the computation of Algorithm~\ref{alg:fastRIScomput}, for the variables $a$, $b$ and $\ell$ the following property holds: 
For any $\ell'\in \{0,\ldots,\ell\}$ and $\ell'$-stable tuple $(x,y)$: $a\ge x$ and $b\ge y$.
\end{propo}

\PF
Consider the initial choices $a=|I\cap V_v|$, $b=|I\cap V_w|$ and $\ell=a+b$, and any $\ell'$-stable tuple $(x,y)$ for $\ell'\le \ell$.
If $x=0$, then obviously $a\ge x$. Otherwise,
$x=\ell'-\RIS_y^I(w)\le \ell-|I\cap V_w|=a$. Analogously, $b\ge y$ follows.

Now suppose that the claim holds for $(a',b')$, and that $(a,b)$ is obtained from this tuple as shown in the algorithm. 
More precisely, $(x,y)$ is an $\ell'$-stable tuple for $\ell'\le \ell$, and
we have $a'\ge x$, $b'\ge y$,  
$a=\max\{0,\ell-\RIS_{b'}^I(w)\}$ and $b=\max\{0,\ell-\RIS_{a'}^I(v)\}$.
We prove that $b\ge y$.
The case $y=0$ is trivial, so now assume $y\ge 1$, and therefore by $\ell'$-stability of $(x,y)$, $y=\ell'-\RIS_x^I(v)$.
Since $a'\ge x$, Proposition~\ref{propo:RISmonotone} yields $\RIS_{a'}^I(v)\le \RIS_x^I(v)$. We conclude that 
$b\ge \ell-\RIS_{a'}^I(v)\ge \ell'-\RIS_x^I(v)=y$.
The inequality $a\ge x$ follows analogously.
It follows that the assignments in Lines~\ref{la:assign1} and~\ref{la:assign2} maintain the invariant.
Finally, decreasing $\ell$ by one (Line~\ref{la:decell}) also obviously maintains the invariant.
\QED

\begin{lem}
\label{lem:RIScomputCorrect}
Algorithm~\ref{alg:fastRIScomput} correctly computes the values $\RIS_{\ell}^I(u)$ for all $\ell\in \{0,\ldots,|I\cap V_u|\}$.
\end{lem}

\PF
The repeat-until loop terminates when $a=\max\{0,\ell-\RIS_b^I(w)\}$ and $b=\max\{0,\ell-\RIS_a^I(v)\}$, so when $(a,b)$ is $\ell$-stable for $I$ and $u$.
By Proposition~\ref{propo:abovestable}, for any $\ell$-stable tuple $(x,y)$ it holds that $a\ge x$ and $b\ge y$, so $(a,b)$ is a maximum $\ell$-stable tuple. Then Lemma~\ref{lem:RISunion} shows that the assignment $\RIS_\ell^I(u)=\RIS_a^I(v)+\RIS_b^I(w)$ is correct.
\QED

To bound the complexity of Algorithm~\ref{alg:fastRIScomput}, we need the following invariant.

\begin{propo}
\label{propo:prestable}
At any time during the computation of Algorithm~\ref{alg:fastRIScomput}, for the variables $a$, $b$ and $\ell$ the following property holds: 
$a\ge \ell-\RIS_b^I(w)$ and $b\ge \ell-\RIS_a^I(v)$. 
\end{propo}

\PF
For the initial choices of $a$, $b$ and $\ell$, the claim holds, since $\ell=a+b$, $\RIS_b^I(w)\ge |I\cap V_w|=b$ and $\RIS_a^I(v)\ge |I\cap V_v|=a$.
Now suppose that the claim holds for $(a',b')$, and that $(a,b)$ is obtained from this tuple as shown in the algorithm. 
More precisely, 
we have $a'\ge \ell-\RIS_{b'}^I(w)$, 
$b'\ge \ell-\RIS_{a'}^I(v)$, 
$a=\max\{0,\ell-\RIS_{b'}^I(w)\}$ and $b=\max\{0,\ell-\RIS_{a'}^I(v)\}$.
We first argue that $a\le a'$. If $a=0$, the statement is clear (using the obvious invariant that these values remain nonnegative). Otherwise, we can write $a=\ell-\RIS_{b'}^I(w)\le a'$. 
By Proposition~\ref{propo:RISmonotone}, it follows that $\RIS_a^I(v)\ge \RIS_{a'}^I(v)$, and therefore
$b\ge \ell-\RIS_{a'}^I(v)\ge \ell-\RIS_a^I(v)$. 
Analogously, $a\ge \ell-\RIS_b^I(w)$ follows.
This shows that the assignments in Lines~\ref{la:assign1} and~\ref{la:assign2} maintain the invariant. Clearly, decreasing $\ell$ by one (Line~\ref{la:decell}) maintains the invariant as well.
\QED

\begin{lem}
\label{lem:RIScomputComplexity}
Algorithm~\ref{alg:fastRIScomput} terminates in time $O(|I\cap V_u|)$. 
\end{lem}

\PF
Denote $n=|I\cap V_u|$.
All lines take constant time (we may assume that the quantities $|I\cap V_v|$ and $|I\cap V_w|$ are known). 
Therefore it suffices to show that in total, the variables $a$ and $b$ are reassigned at most $2n$ times (in Lines~\ref{la:assign1} and~\ref{la:assign2}).
Proposition~\ref{propo:prestable} shows that whenever a new tuple $(a,b)$ is obtained from a previous tuple $(a',b')$, that $a'\ge a$ and $b'\ge b$. 
(If $a=0$, the statement is clear, and otherwise $a=\ell-\RIS_{b'}^I(w)\le a'$ holds. $b\le b'$ follows similarly.)
If $a'=a$ and $b'=b$, then $\ell$ is subsequently decreased (Line~\ref{la:decell}), which is done $n$ times in total. 
Otherwise, $a'+b'>a+b$, and since both values remain nonnegative throughout, this can occur at most $n$ times as well. Hence no line of the algorithm is visited more than $2n$ times.
\QED

\begin{thm}
\label{thm:fastRIScomputation}
Let $T$ be a generalized cotree of a graph $G$ on $n$ vertices, and let $I$ be an \indset\ of $G$. If the values $\RIS_{\ell}^I(v)$ are known for all nontrivial leaves $v\in V(T)$ and all relevant integers $\ell$, then there is an algorithm that computes
\begin{itemize}
 \item the values $\RIS_{\ell}^I(u)$ for all $u\in V(T)$ and $\ell\in \{0,\ldots,|I\cap V_u|\}$, and
 \item the maximum $\ell$-stable tuples for all union nodes $u\in V(T)$ and $\ell\in \{0,\ldots,|I\cap V_u|\}$,
\end{itemize}
with time complexity $O(M)\subseteq O(n^2)$, where $M=\sum_{u\in V(T)} |I\cap V_u|$.
\end{thm}

\PF
The Lemmas~\ref{lem:RIScomputCorrect} and~\ref{lem:RIScomputComplexity} 
show that for a union node $u$, Algorithm~\ref{alg:fastRIScomput} computes the values $\RIS_{\ell}^I(u)$ for all $\ell\in \{0,\ldots,|I\cap V_u|\}$ in time $O(|I\cap V_u|)$, given that the corresponding values for all child nodes are known. 
For the case that $u$ is a trivial leaf or join node, the same claim follows easily from Propositions~\ref{propo:RISleaf} and~\ref{propo:RISjoin}.
So, using a straightforward bottom up computation, all values $\RIS_{\ell}^I(u)$ can be computed correctly in time $O(M)$. 
That is, in constant time on average per entry. It remains to bound $M$ in terms of $n$.

For a node $u\in V(T)$, define $f(u)=\sum_v |V_v|$, where the sum is over all descendants $v$ of $u$ in $T$, including $u$ itself. By induction over $T$, we show that $f(u)\le |V_u|^2$. The induction base is trivial. For the induction step, consider a node $u$ with children $v$ and $w$, and write $a=|V_v|$ and $b=|V_w|$. Then using the induction hypothesis, we can write
\[
f(u)=|V_u|+f(v)+f(w)\le (a+b)+a^2+b^2\le 2ab+a^2+b^2=(a+b)^2=|V_u|^2. 
\]
(We used $a\ge 1$ and $b\ge 1$.)
Let $r$ be the root of $T$. Using
\[ 
M=\sum_{u\in V(T)} |I\cap V_u|\le \sum_{u\in V(T)} |V_u|=f(r)\le n^2,
\]
the statement follows.
\QED

\subsection{Top Down Dynamic Programming Rules}
\label{sec:topdown}

Throughout this section, $T$ denotes again a generalized cotree of $G$ and $I$ denotes an \indset\ of $G$.
In this section, we will show how the values $\freedom^I_k(v)$ can be computed for all nodes $v\in V(T)$. For the case that $v$ is a union node, this requires knowledge of a maximum $\ell$-stable tuple (characterized in Lemma~\ref{lem:UniqueStableCharacterization}).
For the root node of $T$, the value is trivial.

\begin{propo}
\label{propo:TDDP_freedom_root}
Let $r$ be the root node of the cotree $T$. Then
$\freedom^I_k(r)=k$.
\end{propo}

\begin{propo}
\label{propo:TDDP_freedom_join}
Let $u\in V(T)$ be a join node, with children $v$ and $w$ such that $I\cap V_w=\emptyset$.
Then $\freedom^I_k(v)=\freedom^I_k(u)$ and $\freedom^I_k(w)=0$.
\end{propo}

\PF
Considering $I$ itself, $\freedom^I_k(w)=0$ follows immediately.
The inequality $\freedom^I_k(v)\le \freedom^I_k(u)$ follows since $V_v\subseteq V_u$. It remains to prove that $\freedom^I_k(v)\ge \freedom^I_k(u)$.

Consider a {\em shortest} $k$-\TARseq\ $I_0,\ldots,I_p$ in $G$ from $I$ to any \indset\ $I_p$ with $|I_p\cap V_v|=\freedom^I_k(v)$.
If $p=0$, then $\freedom^I_k(u)\le |I\cap V_u|=|I\cap V_v|=\freedom^I_k(v)$, so now assume $p\ge 1$. 
Then $I_p$ is obtained from $I_{p-1}$ by removing a vertex from $V_v$. Since $G_u$ is the complete join of $G_v$ and $G_w$ and $I_{p-1}$ is an \indset, $I_{p-1}\cap V_w=\emptyset$. So $\freedom^I_k(u)\le |I_{p}\cap V_u|=|I_p\cap V_v|=\freedom^I_k(v)$.
\QED

\begin{lem}
\label{lem:TDDP_freedom_union}
Let $u\in V(T)$ be a union node, with left child $v$ and right child $w$. 
Let $\ell=\freedom^I_k(u)$, and let $(x,y)$ be the maximum $\ell$-stable tuple for $I$ and $u$. Then $\freedom^I_k(v)=x$ and $\freedom^I_k(w)=y$.
\end{lem}

\PF
Denote again $I_u=I\cap V(G_u)$.
By Lemma~\ref{lem:UniqueStableCharacterization}, for the maximum $\ell$-stable tuple $(x,y)$ for $I$ and $u$ it holds that 
\begin{equation} 
\label{eq:key}
x=\min |J\cap V_v| \mbox{  and } y=\min |J\cap V_w|, 
\end{equation}
where in both cases the minimum is taken over all \indset s $J$ of $G_u$ with $I_u\tar_{\ell}^{G_u} J$. 

\medskip 
We first use this to show that $\freedom^I_k(v)\ge x$ and $\freedom^I_k(w)\ge y$.
Consider a $k$-\TARseq\ $I_0,\ldots,I_p$ for $G$ with $I_0=I$ and $|I_p\cap V_v|=\freedom^I_k(v)$.
For every $i$, denote $I'_i=I_i\cap V_u$, and consider the sequence $I'_0,\ldots,I'_p$.
By definition of $\ell=\freedom^I_k(u)$, for every $i$ it holds that $|I'_i|\ge \ell$, so this is an $\ell$-\TARseq\ for $G_u$, and thus $I_u\tar_{\ell}^{G_u} I'_p$.
Using~(\ref{eq:key}) it then follows that $\freedom^I_k(v)=|I_p\cap V_v|\ge x$. Analogously, $\freedom^I_k(w)\ge y$ follows.

\medskip 
We will now prove that $\freedom^I_k(v)\le x$ and $\freedom^I_k(w)\le y$.
By~(\ref{eq:key}), there exist \indset s $J_1$ and $J_2$ of $G_u$ with 
$I_u\tar_{\ell}^{G_u} J_1$,
$I_u\tar_{\ell}^{G_u} J_2$,
$|J_1\cap V_v|=x$ and $|J_2\cap V_w|=y$.
By the definition of $\ell=\freedom^I_k(u)$, there exists an \indset\ $B$ of $G$ with $I\tar_k^G B$ and $|B\cap V_u|=\ell$.
We can now apply (module) Lemma~\ref{lem:moduleA} twice, with $V_u$ in the role of module $M$, $I_u$ in the role of $A$, and $J_1$ or $J_2$ respectively in the role of $C$ to conclude that there exist \indset s $D_1$ and $D_2$ of $G$ with 
$I\tar_{k}^{G} D_1$,
$I\tar_{k}^{G} D_2$,  
$|D_1\cap V_v|=x$ and $|D_2\cap V_w|=y$.
Thus $\freedom^I_k(v)\le x$ and $\freedom^I_k(w)\le y$.\QED

\section{Algorithm Summary and Main Theorems}
\label{sec:summary}

In this section, we prove the two main theorems, and summarize how the previous facts and dynamic programming rules can be used to decide efficiently whether $A\tar_k^G B$ for any two given \indset s $A$ and $B$ of a $G$, for any graph that satisfies the properties stated in Theorem~\ref{thm:main_alg}.
First, we prove the theorem that characterizes whether $A\tar_k^G B$, using the previously defined values.

\setcounter{repeatthm}{\value{THM:main_combin}}
\begin{repeatthm}
Let $T$ be a generalized cotree for a graph $G$. 
Let $A$ and $B$ be two \indset s of $G$ of size at least $k$.
Then $A\tar_k^G B$ if and only if
\begin{enumerate}  
 \item
 for all nodes $u\in V(T)$, $\freedom^A_k(u)=\freedom^B_k(u)$, and
 \item
 for all leaves $u\in V(T)$, $(A\cap V_u) \tar_{\ell}^{G_u} (B\cap V_u)$, where $\ell=\freedom^A_k(u)$.
\end{enumerate}
\end{repeatthm}

\PF
We first prove the forward direction. Suppose that $A\tar_k^G B$. Then clearly, for any \indset\ $J$ of $G$, $A\tar_k^G J$ holds if and only if $B\tar_k^G J$. So $\freedom_k^A(u)=\freedom_k^B(u)$ holds for every node $u\in V(T)$ (Definition~\ref{def:freedom}), which proves the first property. 

For any $u\in V(T)$, we may now denote $\freedom_k(u)=\freedom_k^A(u)=\freedom_k^B(u)$.
Consider a $k$-\TARseq\ $I_0,\ldots,I_p$ for $G$ from $A$ to $B$. For any node $u\in V(T)$ and any $i\in \{0,\ldots,p\}$, it holds that $|I_i\cap V_u|\ge \freedom_k(u)$ (Definition~\ref{def:freedom}). So $I'_0,\ldots,I'_p$ with $I'_i=I_i\cap V_u$ for all $i$ is a $\freedom_k(u)$-\TARseq\ for $G_u$. This shows that $(A\cap V_u) \tar_{\freedom_k(u)}^{G_u} (B\cap V_u)$, and thus proves the second property.

\medskip
Now we prove the other direction. Assume that the two properties hold. So we may denote $\freedom_k(u)=\freedom_k^A(u)=\freedom_k^B(u)$ for all nodes $u$. 
We prove the following claim by induction over $T$:

\medskip
\noindent
{\em Claim~A:}
For all nodes $u\in V(T)$: $(A\cap V_u)\tar_{\freedom_k(u)}^{G_u} (B\cap V_u)$.
\medskip

{\em Induction base:} For leaf nodes $u\in V(T)$, the statement follows immediately from the second property. 
\medskip 

{\em Induction step:} First consider a join node $u\in V(T)$ with children $v$ and $w$.
Suppose that $\freedom_k(v)\ge 1$. This implies $A\cap V_v\not=\emptyset$ and $B\cap V_v\not=\emptyset$. Therefore, since $u$ is a join node, $A\cap V_u=A\cap V_v$ and $B\cap V_u=B\cap V_v$. In addition, $\freedom_k(u)=\freedom_k(v)$ (Proposition~\ref{propo:TDDP_freedom_join}). 
From these facts, and the induction assumption $(A\cap V_v)\tar_{\freedom_k(v)}^{G_v} (B\cap V_v)$, we conclude that $(A\cap V_u)\tar_{\freedom_k(u)}^{G_u} (B\cap V_u)$. 
The case $\freedom_k(w)\ge 1$ is analog.
Now suppose that $\freedom_k(v)=\freedom_k(w)=0$. Then $\freedom_k(u)=0$ (Proposition~\ref{propo:TDDP_freedom_join}). The desired claim follows since $(A\cap V_u)\tar_{0}^{G_u} (B\cap V_u)$ trivially holds.

\medskip
Next, consider the case that $u\in V(T)$ is a union node with left child $v$ and right child $w$.
Denote $\ell=\freedom_k(u)$, $x=\freedom_k(v)$ and $y=\freedom_k(w)$. By Lemma~\ref{lem:TDDP_freedom_union}, $(x,y)$ is the maximum $\ell$-stable tuple for $u$, for both $A$ and $B$.
We 
define $C_v$ to be an \indset\ of $G_v$ with $(A\cap V_v)\tar_x^{G_v} C_v$, with maximum size among all such sets, and 
define $C_w$ to be an \indset\ of $G_w$ with $(A\cap V_w)\tar_y^{G_w} C_w$, with maximum size among all such sets.
By induction, $(A\cap V_v)\tar_x^{G_v} (B\cap V_v)$, so it also holds that $(B\cap V_v)\tar_x^{G_v} C_v$, and that $C_v$ has maximum size among all such reachable sets. Analogously, $(B\cap V_w)\tar_y^{G_w} C_w$, and $C_w$ has maximum size among all such reachable sets.
Define $C_u=C_v\cup C_w$. We will now show that $C_u$ is reachable from both $A\cap V_u$ and $B\cap V_u$, which proves Claim~A for node $u$.

Lemma~\ref{lem:UniqueStableCharacterization} shows that there exists an \indset\ $J$ of $G_u$ with $(A\cap V_u)\tar_{\ell}^{G_u} J$ and $|J\cap V_v|=x$.
Using this, we argue that there exists an \indset\ $J_1$ of $G_u$ with $(A\cap V_u)\tar_{\ell}^{G_u} J_1$ and $J_1\cap V_v=C_v$.
If $A\cap V_u=\emptyset$, then this claim is trivial. Otherwise, we can apply (module) Lemma~\ref{lem:moduleA} 
to draw this conclusion (using $V_v$, $G_u$, $J$ and $C_v$ in the roles of the module $M$, entire graph $G$, and \indset s $B$ and $C$, respectively).
Analogously, we may conclude that there exists an \indset\ $J_2$ of $G_u$ with $(A\cap V_u)\tar_{\ell}^{G_u} J_2$ and $J_2\cap V_w=C_w$.
Since $C_u=C_v\cup C_w$, we can now apply (module) Lemma~\ref{lem:moduleB} 
(with $G_u$ in the role of the entire graph, $V_v$ and $V_w$ in the roles of disjoint modules $M_1$ and $M_2$, and $J_1$ and $J_2$ in the roles of $B_1$ and $B_2$), 
to conclude that $A\tar_{\ell}^{G_u} C_u$. 
For this, we require the fact that $C_v$ has maximum size among all \indset s of $G_v$ that are reachable from $A\cap V_v$.

The argument from the previous paragraph also holds when replacing $A$ by $B$, 
since $C_v$ and $C_w$ are also maximum reachable \indset s from $B\cap V_v$ and $B\cap V_w$.
Thus we may also conclude that $B\tar_{\ell}^{G_u} C_u$. 
Using the fact that $\tar_{\ell}^{G_u}$ is an equivalence relation, we conclude that $A\tar_{\ell}^{G_u} B$, which proves the desired claim for $u$.

\medskip
This concludes the induction proof of Claim~A. Applying Claim~A to the root node $r$ of $T$ shows that $A\tar_k^G B$, since $\freedom_k(r)=k$ (Proposition~\ref{propo:TDDP_freedom_root}), and $G=G_r$, and therefore concludes the proof of the theorem. 
\QED

Next, we prove our main algorithmic result. In the next section, we give examples of graph classes for which this theorem yields efficient algorithms.

\setcounter{repeatthm}{\value{THM:main_alg}}
\begin{repeatthm}
Let $T$ be a generalized cotree for a graph $G$ on $n$ vertices, let $k\in \mathbb{N}$ and let $A$ and $B$ be \indset s of $G$. 
If for every nontrivial leaf $v\in V(T)$ and relevant integer $\ell$, 
\begin{itemize}
\item 
the values $\RIS^A_\ell(v)$ and $\RIS^B_\ell(v)$ are known, and 
\item 
it is known whether $(A\cap V_v)\tar_\ell^{G_v} (B\cap V_v)$, 
\end{itemize}
then in time $O(n^2)$ it can be decided whether $A\tar_k^G B$.
\end{repeatthm}

\PF
We may assume that $|A|\ge k$ and $|B|\ge k$, otherwise we can immediately answer NO.
First we use a {\em bottom up} dynamic programming algorithm, to compute the values $\RIS^A_{\ell}(u)$ and $\RIS^B_{\ell}(u)$ for every node $u$ and relevant integer $\ell$. Theorem~\ref{thm:fastRIScomputation} shows that this can be done in time $O(n^2)$, and that at the same time the maximum $\ell$-stable tuples can be computed for $A$, $B$, all union nodes $u$ and relevant integers $\ell$. 
(Recall that this uses the dynamic programming rules for trivial leaves, join nodes and union nodes given in Proposition~\ref{propo:RISleaf}, Proposition~\ref{propo:RISjoin} and Lemma~\ref{lem:RISunion}, respectively, and the fast computation of maximum $\ell$-stable tuples given in Section~\ref{ssec:fastRIScomput}.)

\medskip
Next, we start the {\em top down} phase of the dynamic programming algorithm, where we compute the values $\freedom_k^A(u)$ and $\freedom_k^B(u)$ for every node $u$.
For the root node $r$ of $T$, we can initialize these values to $k$ (Proposition~\ref{propo:TDDP_freedom_root}). Next, for every node $u$ for which these two values are known, we can compute these two values for the two children $v$ and $w$, by applying Proposition~\ref{propo:TDDP_freedom_join} for join nodes and Lemma~\ref{lem:TDDP_freedom_union} for union nodes. Note that applying Lemma~\ref{lem:TDDP_freedom_union} to a union node $u$ requires the previously computed maximum $\ell$-stable tuple $(x,y)$ for $I=A,B$, with $\ell=\freedom_k^I(u)$. This is why the bottom up phase is required.

\medskip 

Finally, we return YES if
\begin{itemize}
 \item
 for all nodes $v\in V(T)$, $\freedom_k^A(v)=\freedom_k^B(v)$, and
 \item
 for all leaves $v\in V(T)$, $(A\cap V_v) \tar_{\ell}^{G_v} (B\cap V_v)$, where $\ell=\freedom_k^A(v)$.
\end{itemize}
This is correct by Theorem~\ref{thm:NEW_main_combin}.
(Note that for trivial leaves $v\in V(T)$, $(A\cap V_v) \tar_{\ell}^{G_v} (B\cap V_v)$ always holds, and for nontrivial leaves, we assume that this information is given.)
Considering the dynamic programming rules, every value that is assigned in the top down phase can be computed in constant time per value. Hence the top down phase takes time $O(|V(T)|)=O(n)$, and the total complexity of the algorithm becomes $O(n^2)$ (which is dominated by the bottom up phase).
\QED

\section{Graph Classes}
\label{sec:graphclasses}

In this section, we discuss graph classes to which Theorem~\ref{thm:main_alg} applies.
Firstly, Theorem~\ref{thm:main_alg} easily implies that the TAR-Reachability problem can be decided efficiently on cographs.

\begin{thm}
\label{thm:cograph}
Let $G$ be a cograph on $n$ vertices, let $k\in \mathbb{N}$ and let $A$ and $B$ be \indset s of $G$. In time $O(n^2)$ it can be decided whether $A\tar_k^G B$.
\end{thm}

\PF
A cotree $T$ for $G$ can be constructed in linear time~\cite{CPS85}. 
We can easily guarantee that this is a binary tree.
Since a cotree only has trivial leaves, Theorem~\ref{thm:main_alg} can now be applied. 
Indeed, for a trivial leaf $u$: Proposition~\ref{propo:RISleaf} shows that $\RIS^I_\ell(u)=1$ holds for all relevant $\ell\in \{0,1\}$. 
Secondly, it can be seen that $(A\cap V_v)\tar_\ell^{G_v} (B\cap V_v)$ always holds for all relevant $\ell\in \{0,1\}$. So the conditions of Theorem~\ref{thm:main_alg} are satisfied.
\QED

Combining this theorem with Lemma~\ref{lem:TJisTAR} shows that we can efficiently decide whether $A\tj^G B$ in the case that $G$ is a cograph, which answers an open question from~\cite{KMM12}:

\begin{corol}
Let $G$ be a cograph on $n$ vertices, and let $A$ and $B$ be \indset s of $G$. 
In time $O(n^2)$ it can be decided whether $A\tj^G B$.
\end{corol}

Theorem~\ref{thm:main_alg} is however much stronger, and implies that TAR-Reachability can be decided efficiently for much richer graph classes.
We will now give an example of such a graph class, namely the class of {\em all graphs that admit a cotree decomposition into chordal graphs.} Along the way, we will introduce some tools that allow proving the same for other graph classes that can be obtained by taking unions and joins of graphs from a graph class $\GG$, for which the values/properties from Theorem~\ref{thm:main_alg} can efficiently be computed/decided.

A graph $G$ is {\em chordal} if it contains no cycles of length four or more as induced subgraphs (in other words: if every cycle of length at least four contains a {\em chord}).
The only two properties of chordal graphs that we will use are the following. Firstly:

\begin{thm}[\cite{Gav72}]
\label{thm:ChordalMaxIndSet}
Let $G$ be a chordal graph. Then $\alpha(G)$ can be computed in polynomial time. 
\end{thm}

This statement is well-known, and relatively easy to prove using the concept of simplicial vertices. For the more general class of {\em perfect graphs}, $\alpha(G)$ can in fact also be computed in polynomial time. See~\cite[Section~66.3]{Schrijver} for more background.
Secondly, we use the fact that chordal graphs are obviously even-hole-free. A graph $G$ is {\em even-hole-free} if it contains no even cycles as induced subgraphs. To our knowledge, no polynomial time algorithm for computing $\alpha(G)$ for even-hole-free graphs is known; otherwise, the result from this section could be generalized to even-hole-free graphs. See also~\cite{Vus10,KMM12}.
We will also apply the following result, which was proved in~\cite{KMM12}.

\begin{thm}[\cite{KMM12}]
\label{thm:EvenHoleFreeTJ}
Let $A$ and $B$ be two \indset s of an even-hole-free graph $G$ with $|A|=|B|$. Then $A\tj^G B$. 
\end{thm}

Using Lemma~\ref{lem:TJisTAR}, this theorem can be applied to the TAR model to conclude:

\begin{lem}
\label{lem:EvenHoleFreeTAR}
Let $A$ and $B$ be two distinct \indset s of an even-hole-free graph $G$.
Then $A\tar_k^G B$ if and only if
neither $A$ nor $B$ is a dominating set of size $k$.
\end{lem}

\PF
If $A$ is a dominating set of size $k$, then no token can be added to $A$, and no token can be removed from $A$. So $A$ has no neighbors in $\TAR_k(G)$. Since $A$ and $B$ are distinct, it follows that $A\not\tar_k^G B$. This follows similarly if $B$ is a dominating set of size $k$.

Now suppose that neither $A$ nor $B$ is a dominating set of size $k$. Then we show that $A\tar_k^G B$. We can easily construct an \indset\ $A'$ with $A\tar_k^G A'$ and $|A'|=k+1$:
\begin{itemize}
 \item If $|A|=k$ then add an arbitrary vertex $v$ which has no neighbors in $A$ (which exists since $A$ is not dominating).
 \item If $|A|\ge k+1$ then remove arbitrary vertices from $A$ until an \indset\ of size $k+1$ is obtained.
\end{itemize}
Similarly, we can easily construct an \indset\ $B'$ with $B\tar_k^G B'$ and $|B'|=k+1$.
By Theorem~\ref{thm:EvenHoleFreeTJ}, $A'\tj^G B'$. Next, Lemma~\ref{lem:TJisTAR} shows that $A'\tar_k^G B'$.
Combining this with $A\tar_k^G A'$ and $B\tar_k^G B'$ shows that $A\tar_k^G B$.\QED

The above lemma easily yields the following statement.

\begin{corol}
\label{cor:EvenHoleFreePreRIS}
Let $I$ be an \indset\ of an even-hole-free graph $G$, and let $J$ be an \indset\ of $G$ with $I\tar_k J$ that maximizes $|J|$ among all such sets. Then
\begin{itemize}
 \item $|J|=|I|$ if $I$ is a dominating set of size $k$, and
 \item $|J|=\alpha(G)$ otherwise.
\end{itemize}
\end{corol}

This in turn gives us immediately an easy way to compute the values $\RIS_\ell^I(u)$ for the case that $G_u$ is even-hole-free:

\begin{corol}
\label{cor:EvenHoleFreeRIS}
Let $T$ be a cotree decomposition of a graph $G$ into even-hole-free graphs, and let $I$ be an \indset\ of $G$. Then for every leaf $u\in V(T)$, and every relevant value $\ell$:
\begin{itemize}
 \item $\RIS_\ell^I(u)=|I|$ if $I\cap V_u$ is a dominating set of $G_u$ of size $\ell$, and
 \item $\RIS_{\ell}^I(u)=\alpha(G_u)$ otherwise.
\end{itemize}
\end{corol}

Combining Theorem~\ref{thm:ChordalMaxIndSet}, Lemma~\ref{lem:EvenHoleFreeTAR} and Corollary~\ref{cor:EvenHoleFreeRIS} shows that if we have a cotree decomposition of a graph $G$ into chordal graphs, then the conditions of Theorem~\ref{thm:main_alg} are satisfied, so we can compute in polynomial time whether $A\tar_k B$. However, it remains to discuss how in general, a cotree decomposition into chordal graphs can be found.
Recall that for a graph $G$, by $\overline{G}$ the {\em complement} of $G$ is denoted, which is the graph $\overline{G}=(V(G),\{uv \mid uv\not\in E(G)\})$.

\begin{defi}
\label{def:indecomp}
A graph $H$ is {\em indecomposable} if both $H$ and $\overline{H}$ are connected.
A {\em maximal cotree decomposition} of a graph $G$ is a generalized cotree decomposition $T$ such that for every leaf $u\in V(T)$, $G_u$ is indecomposable.
\end{defi}

\begin{propo}
\label{propo:MaximalCotreeDecompPolyTime}
For any graph $G$, a maximal cotree decomposition of $G$ can be computed in polynomial time.
\end{propo}

\PF
A polynomial time algorithm for testing whether a given graph is indecomposable follows immediately from the definition (quadratic time in fact). Now consider the following algorithm for constructing a maximal cotree decomposition of $G$: start with a trivial generalized cotree decomposition $T$, consisting of one (root) node $r$ with $G_r=G$. As long as the current generalized cotree decomposition $T$ contains a leaf $u\in V(T)$ such that $G_u$ is decomposable, 
partition the vertices of $G_u$ into new sets $V_v$ and $V_w$ such that $G_u$ is the disjoint union or complete join of $G[V_v]$ and $G[V_w]$ (this can be trivially done in the case where $G_u$ is disconnected, respectively in the case where $\overline{G_u}$ is disconnected). Now add corresponding new leaf nodes $v$ and $w$ as children of $u$, and make $u$ into a union or join node, respectively. This way, a generalized cotree decomposition of $G$ is maintained. The algorithm terminates after at most $|V(G)|$ steps (which all take polynomial time), since in every step, the number of leaves of $T$ increases by one, 
and a generalized cotree decomposition has at most $|V(G)|$ leaves. When the algorithm terminates, the resulting generalized cotree decomposition is clearly maximal.
\QED

A graph class $\GG$ is called {\em hereditary} if for every $G\in \GG$ and every induced subgraph $H$ of $G$, $H\in \GG$ holds. Clearly, chordal graphs are hereditary.

\begin{lem}
\label{lem:AnyMaximalDecompSuffices}
Let $\GG$ be a hereditary graph class, and let $G$ be a graph that admits a cotree decomposition into $\GG$-graphs. Then every maximal cotree decomposition of $G$ is a cotree decomposition into $\GG$-graphs.
\end{lem}

\PF 
Let $T^*$ be a maximal cotree decomposition of $G$, and let $T^C$ be a cotree decomposition of $G$ into $\GG$-graphs.
Denote by $G^*_u$ and $G^C_u$ the subgraphs of $G$ that correspond to nodes $u\in V(T^*)$ and $u\in V(T^C)$, respectively. Similarly, denote their vertex sets by $V^*_u$ and $V^C_u$.

Consider a leaf $u\in V(T^*)$. We will prove that $G_u^*$ is also part of the graph class $\GG$.
If there is a leaf node $v\in V(T^C)$ such that $V^*_u\subseteq V^C_v$, then $G^*_u$ is an induced subgraph of $G^C_v$, so since $\GG$ is hereditary, $G^*_u\in \GG$ holds.

Now assume that there is no such leaf node in $T^C$. Then observe that we may consider a node $w\in V(T^C)$ with $V^*_u\subseteq V^C_w$, which has no child nodes that satisfy this property. 
(In other words: $w$ is a lowest common ancestor of all nodes $x$ with $V^C_x\cap V^*_u\not=\emptyset$.)
Let $x$ and $y$ be the two child nodes of $w$.
So by choice of $w$, $V^*_u$ can be partitioned into two nonempty sets $S=V^*_u\cap V^C_x$ and $T=V^*_u\cap V^C_y$.
If $w$ is a join node, then $G^*_u$ can be written as the complete join of $G[S]$ and $G[T]$, so $\overline{G^*}$ is disconnected, contradicting the fact that it is indecomposable. Similarly, if $w$ is a union node, then $G^*_u$ can be written as the disjoin union of $G[S]$ and $G[T]$, so it is disconnected, contradicting the fact that it is indecomposable. This concludes the proof that for every $u\in V(T^*)$, $G_u\in \GG$ holds.
\QED

We now summarize how the previous statements yield a polynomial time algorithm for testing $A\tar_k^G B$, whenever $G$ is a graph that admits a cotree decomposition into chordal graphs.

\begin{thm}
\label{thm:TAR_CotreeIntoChordal}
Let $G$ be a graph that admits a cotree decomposition into chordal graphs, and let $A$ and $B$ be \indset s of $G$, both of size at least $k$.
Then in polynomial time, we can decide whether $A\tar_k^G B$.
\end{thm}

\PF
We first construct a maximal cotree decomposition $T$ of $G$ in polynomial time (Proposition~\ref{propo:MaximalCotreeDecompPolyTime}). By Lemma~\ref{lem:AnyMaximalDecompSuffices}, $T$ is then a cotree decomposition into chordal graphs (since chordal graphs are hereditary).
So by Theorem~\ref{thm:ChordalMaxIndSet}, we can compute $\alpha(G_u)$ for every leaf $u\in V(T)$.
Combining this with Corollary~\ref{cor:EvenHoleFreeRIS}, and the fact that chordal graphs are even-hole-free, shows that we can compute the values $\RIS_\ell^A(u)$ and $\RIS_\ell^B(u)$ for every leaf $u\in V(T)$ and relevant $\ell$.
Finally, Lemma~\ref{lem:EvenHoleFreeTAR} gives an easy way to decide in polynomial time whether $A\tar_{\ell}^{G_u} B$ for any leaf $u\in V(T)$ and relevant value $\ell$. So the conditions of Theorem~\ref{thm:main_alg} are satisfied for the generalized cotree decomposition $T$ of $G$, and thus we can compute in polynomial time whether $A\tar_k^G B$.
\QED

\section{A Linear Bound on the Diameter of the Solution Graph}
\label{sec:diameter}

Using the previous lemmas, we can efficiently decide whether there exists a $k$-\TARseq\ from $A$ to $B$ in a cograph $G$. However, from these lemmas, one cannot easily deduce a polynomial upper bound for the length of such a sequence. This requires studying the aforementioned cascading sequences in more detail, which is what we will do in this section. We will show that if $A\tar_k^G B$, then there exists a $k$-\TARseq\ from $A$ to $B$ of length at most $4n-|A|-|B|$, where $n=|V(G)|$.

The main idea is as follows. Given \indset s $A$ and $B$ of $G$ with $A\tar_k^G B$, we choose an appropriate subgraph $G'$ of $G$ such that there exist maximum \indset s $A'$ and $B'$ of $G$ and short $k$-\TARseq s from $A$ to $A'$ and from $B$ to $B'$. These sequences are short in the sense that for every vertex $v\in V(G)$, no token is added on $v$ after the first token is removed from $v$. So in total, there are at most $2n-|A|-|A'|$ token additions/removals used in the sequence from $A$ to $A'$, and a similar statement holds for $B$ and $B'$. Finally, we show that $k$-\TARseq\ from $A'$ to $B'$ exists, of length at most $|A\Delta B|$.
(Recall that $A\Delta B=(A\bs B)\cup (B\bs A)$ denotes the {\em symmetric difference} of $A$ and $B$.) Combining these three $k$-\TARseq s yields a $k$-\TARseq\ from $A$ to $B$ of length at most $4n-|A|-|B|$ in the subgraph $G'$, and therefore also in $G$.

\begin{figure}
\centering
\scalebox{1}{$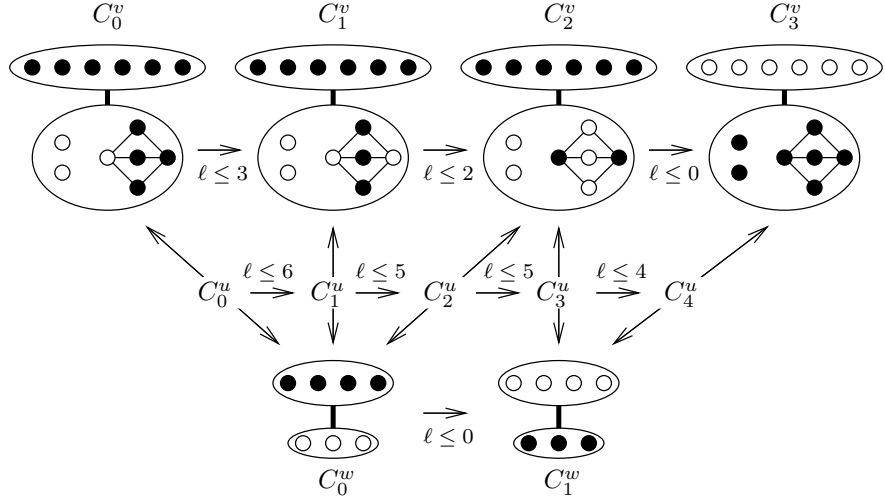$}
\caption{SU-sequences $C^v_0,\ldots,C^v_3$ and $C^w_0,C^w_1$ for the graphs $G_v$ and $G_w$ from Figure~\ref{fig:cascading}, and a resulting SU-sequence $C^u_0,\ldots,C^u_4$ for their disjoint union $G_u$.}
\label{fig:SUseqs}
\end{figure}

We now define the type of \TARseq s that we will consider. For a node $u\in V(T)$ and {\em every} value of $\ell\in \{0,\ldots,|I\cap V_u|\}$, a {\em subsequence} of the next sequence outlines an $\ell$-\TARseq\ for $G_u$ from $I\cap V_u$ to an \indset\ $J$ with $|J|=\RIS_\ell^I(u)$ (Properties~\ref{it:begin} and~\ref{it:shortseq}). In addition it is {\em short} in the sense that every vertex of $G_u$ is used for at most one token addition (Property~\ref{it:newverts}). This motivates the name {\em short universal sequence}. Examples of these sequences are given in Figure~\ref{fig:SUseqs} for the graphs $G_v$ and $G_w$ from Figure~\ref{fig:cascading}.

\begin{defi}
\label{def:SUseq}
Let $T$ be a cotree of a graph $G$, and let $I$ be an \indset\ of $G$. 
A {\em short universal sequence} or {\em SU-sequence} for a node $u\in V(T)$, based on $I$, is a sequence 
$C_0,\ldots,C_{p}$ of \indset s of $G_u$ that satisfy the following properties:
\begin{enumerate}
 \item
 \label{it:begin}
 $C_0=I\cap V_u$.
 \item
 \label{it:increase}
 For all $i\in\{0,\ldots,p-1\}$: $|C_{i+1}|>|C_i|$.
 \item
 \label{it:newverts}
 For all $i,j\in\{0,\ldots,p-1\}$ with $i\le j$: $(C_{j+1}\bs C_j)\cap C_i=\emptyset$. 
 \item
 \label{it:shortseq}
 For all $\ell\in \{0,\ldots,|I\cap V_u|\}$ and $i\in \{0,\ldots,p\}$: 
 if $|C_i|<\RIS^I_{\ell}(u)$ then $i<p$ and there exists an $\ell$-\TARseq\ in $G_u$ from $C_i$ to $C_{i+1}$ that only adds tokens on $C_{i+1}\bs C_i$ 
\end{enumerate}
\end{defi}

Below we will show by induction over $T$ that for every node $u\in V(T)$, an SU-sequence exists. But first, we will prove two properties that indicate why these sequences are useful for finding short $\ell$-\TARseq s from $I\cap V_u$ to an \indset\ $J$ with $|J|=\RIS^I_{\ell}(u)$, for any value of $\ell$.

\begin{propo}
\label{propo:SU_ends_with_maxindset}
Let $T$ be a cotree of a graph $G$, and let $I$ be an \indset\ of $G$. 
Let $C_0,\ldots,C_{p}$ be an SU-sequence for a node $u\in V(T)$, based on $I$.
Then $|C_p|=\alpha(G_u)$.
\end{propo}
\PF
For all $i\in \{0,\ldots,p\}$, since $C_i$ is an \indset\ of $G_u$, $|C_i|\le \alpha(G_u)$. If the inequality is strict, then $|C^u_i|<\alpha(G_u)=\RIS^I_0(u)$, so Property~\ref{it:shortseq} of Definition~\ref{def:SUseq} shows that $i<p$.
\QED

\begin{lem}
\label{lem:SUseq_usage}
Let $T$ be a cotree of a graph $G$, and let $I$ be an \indset\ of $G$. 
Let $C_0,\ldots,C_{p}$ be an SU-sequence for a node $u\in V(T)$, based on $I$.
For any $\ell\in \{0,\ldots,|I\cap V_u|\}$ and $i\in \{0,\ldots,p-1\}$ with 
$|C_i|<\RIS^I_{\ell}(u)$, there exists an $\ell$-\TARseq\ from $I\cap V_u$ to $C_{i+1}$ in $G_u$ of length $2(|\bigcup_{j=0}^{i+1} C_j|)-|C_0|-|C_{i+1}|$. 
Therefore, $\RIS_{\ell}^I(u)\ge |C_{i+1}|$.
\end{lem}

\PF
For any $\ell$, we prove the statement by induction over $i$. 
We will assume that $|C_i|<\RIS^I_{\ell}(u)$, otherwise there is nothing to prove.

For $i=0$, Property~\ref{it:shortseq} shows that there is an $\ell$-\TARseq\ in $G_u$ from $C_0$ to $C_1$ that only adds tokens on $C_{1}\bs C_0$. It follows that this \TARseq\ uses exactly $|C_{1}\bs C_0|$ token additions, and $|C_0\bs C_{1}|$ token removals. We can write
\[
|C_1\bs C_0|+|C_0\bs C_1|=(|C_0\cup C_1|-|C_0|)+(|C_0\cup C_1|-|C_1|),
\]
which proves the statement.

Now suppose $i\ge 1$. Since $|C_i|<\RIS^I_{\ell}(u)$, Property~\ref{it:increase} shows that $|C_{i-1}|<\RIS^I_{\ell}(u)$ holds as well. So by induction, there exists an $\ell$-\TARseq\ in $G_u$ from $C_0$ to $C_i$ of length 
$2(|\bigcup_{j=0}^{i} C_j|)-|C_0|-|C_{i}|$. 
By Property~\ref{it:shortseq}, there also exists an $\ell$-\TARseq\ in $G_u$ from $C_i$ to $C_{i+1}$, of length $|C_i\bs C_{i+1}|+|C_{i+1}\bs C_i|$ (using an argument similar to above). These can be combined into an $\ell$-\TARseq\ in $G_u$ from $C_0$ to $C_{i+1}$.
The total length of this sequence is therefore:
\[
2\bigg|\bigcup_{j=0}^i C_j\bigg| - |C_0| - |C_i| + |C_i\bs C_{i+1}| + |C_{i+1}\bs C_i| =  
\]
\[
2\bigg|\bigcup_{j=0}^i C_j\bigg| - |C_0| - |C_i| + 2|C_i\cup C_{i+1}|-|C_i|-|C_{i+1}| =  
\]
\[
2\bigg|\bigcup_{j=0}^i C_j\bigg| + 2|C_{i+1}\bs C_i| - |C_0| - |C_{i+1}|\stackrel{\mbox{\footnotesize{(Property~\ref{it:newverts})}}}{=}  
\]
\[
2\bigg|\bigcup_{j=0}^i C_j\bigg| + 2\bigg|C_{i+1}\bs \bigg(\bigcup_{j=0}^i C_j\bigg)\bigg| - |C_0| - |C_{i+1}| =  
\]
\[
2\bigg|\bigcup_{j=0}^{i+1} C_j\bigg| - |C_0| - |C_{i+1}|.
\]

This concludes the induction proof, so we conclude that for any $i$ with $|C_i|<\RIS^I_{\ell}(u)$, there exists an $\ell$-\TARseq\ from $I\cap V_u$ to $C_{i+1}$ in $G_u$ of length $2(|\bigcup_{j=0}^{i+1} C_j|)-|C_0|-|C_{i+1}|$. From this, it follows immediately that $\RIS_{\ell}^I(u)\ge |C_{i+1}|$.
\QED

We will now prove that SU-sequences exist for every $u\in V(T)$. 
\begin{propo}
\label{propo:SUseq_leaf}
Let $T$ be a cotree of a graph $G$, and let $I$ be an \indset\ of $G$. 
For every leaf node $u\in V(T)$, there exists an SU-sequence based on $I$.
\end{propo}
\PF
We define $C_0=I\cap V_u$ and $C_{p}=\{u\}$.
Choose $p=0$ if these two sets are the same, and $p=1$ otherwise.
One can easily verify that the four properties from Definition~\ref{def:SUseq} hold for this sequence. (Recall that by Proposition~\ref{propo:RISleaf}, $\RIS^I_{\ell}(u)=1$ for all $\ell$.)
\QED

\begin{lem}
\label{lem:SUseq_join} 
Let $T$ be a cotree of a graph $G$, and let $I$ be an \indset\ of $G$. 
Let $u\in V(T)$ be a join node with children $v$ and $w$. 
If there exist SU-sequences for $v$ and $w$, then there exists an SU-sequence for $u$ (all based on $I$).
\end{lem}

\PF
Suppose $u$ is a join node, with children $v$ and $w$.
We will construct an SU-sequence $C^u_0,\ldots,C^u_{p^u}$ for $u$.

If $I\cap V_u=\emptyset$, then we choose $p^u=1$, $C^u_0=\emptyset$ and $C^u_1$ to be a maximum \indset\ of $G_u$. This choice satisfies the four properties of Definition~\ref{def:SUseq}. 
So now we may assume w.l.o.g. that $I\cap V_v\not=\emptyset$ and $I\cap V_w=\emptyset$.
By induction, for $v$ there exists an SU-sequence $C^v_0,\ldots,C^v_{p^v}$ based on $I$. 
For all $i\in \{0,\ldots,p^v\}$, we choose $C^u_i=C^v_i$. 
If $C^v_i$ is also a maximum \indset\ of $G_u$ then this is the entire sequence for $u$ (so we choose $p^u=p^v$), and it satisfies the four properties again 
(recall that in this case, $\RIS^I_{\ell}(u)=\RIS^I_{\ell}(v)$ for all $\ell$, by Proposition~\ref{propo:RISjoin}). 
Otherwise, since $u$ is a join node, any set $J\subseteq V_u$ is a maximum \indset\ for $G_u$ if and only if it is a maximum \indset\ for $G_v$. 
Therefore we can choose $p^u=p^v+1$, and choose $C^u_{p^u}$ to be any maximum independent set of $G_v$. 
Clearly, there exists a $0$-\TARseq\ from 
$C^u_{p^u-1}=C^v_{p^v}$ to $C^u_{p^u}$ that only adds tokens on $C^u_{p^u}\bs C^u_{p^u-1}$.
Since $\RIS^I_0(u)=|C^u_{p^u}|$ and $\RIS^I_{\ell}(u)=\RIS^I_{\ell}(v)$ for all $\ell\ge 1$ (Proposition~\ref{propo:RISjoin}), this shows that Property~\ref{it:shortseq} again holds for the new sequence.
For all $i<p^u$, $C^u_i\subseteq V_v$, so $C^u_{p^u}\cap C^u_i=\emptyset$, and thus Property~\ref{it:newverts} is again satisfied for this new sequence.
Property~\ref{it:begin} holds since $C^u_0=|I\cap V_v|=|I\cap V_u|$ (using induction, and that $u$ is a join node with $I\cap V_v\not=\emptyset$, respectively). 
\QED

The proof of the following lemma is also illustrated in Figure~\ref{fig:SUseqs}. Given SU-sequences for children $v$ and $w$ of a union node $u$, we obtain an SU-sequence for $u$ by letting every set in the new sequence be the union of one set from the SU-sequence for $v$ and one set from the SU-sequence for $w$.

\begin{lem}
\label{lem:SUseq_union} 
Let $T$ be a cotree of a graph $G$, and let $I$ be an \indset\ of $G$. 
Let $u\in V(T)$ be a union node with children $v$ and $w$. 
If there exist SU-sequences for $v$ and $w$, then there exists an SU-sequence for $u$ (all based on $I$).
\end{lem}

\PF
Let $C^v_0,\ldots,C^v_{p^v}$ and $C^w_0,\ldots,C^w_{p^w}$ be SU-sequences based on $I$ for $v$ and $w$, respectively.
We will construct a SU-sequence $C^u_0,\ldots,C^u_{p^u}$ for $u$ from these. 
These sets will be constructed such that for every index $a$, there exist indices $b$ and $c$ with
\[
C^u_a=C^v_b\cup C^w_c.
\]
First we choose $C^u_0=C^v_0\cup C^w_0$, which guarantees that $C^u_0=I\cap V_u$, so Property~\ref{it:begin} is satisfied. 
Next, for every choice of indices $a,b,c$ such that we assigned $C^u_a=C^v_b\cup C^w_c$, continue the construction of the sequence according to the following method.
For notational convenience, we define $\RIS^I_i(v)=\RIS^I_0(v)$ and $\RIS^I_i(w)=\RIS^I_0(w)$ for all $i<0$.
\begin{enumerate}[(a)]
\item 
Denote $q=|C^v_b|$ and $r=|C^w_c|$.
\item
\label{it:seqend}
If $q=\alpha(G_v)$ and $r=\alpha(G_w)$ then assign $p^u:=a$ (so the SU-sequence for $u$ ends here). 
Otherwise, choose $C^u_{a+1}$ as follows:
\item 
\label{it:ellchoice}
Choose $\ell$ to be the maximum value in $\{0,\ldots,|I\cap V_u|\}$ such that 
$\RIS^I_{\ell-r}(v)>q$ or
$\RIS^I_{\ell-q}(w)>r$.
\item 
If $\RIS^I_{\ell-r}(v)>q$ then choose $C^u_{a+1}=C^v_{b+1}\cup C^w_{c}$, and otherwise (when $\RIS^I_{\ell-q}(w)>r$) choose $C^u_{a+1}=C^v_b\cup C^w_{c+1}$.
\end{enumerate}
We first argue that a value $\ell$ can always be chosen as in~(\ref{it:ellchoice}): 
If $q=|C^v_b|<\alpha(G_v)$, then by Property~\ref{it:shortseq}, $\RIS^I_{0}(v)>q$. So choosing any $\ell\in\{0,\ldots,|I\cap V_u|\}$ with $\ell\le r$ suffices. 
Otherwise, by~(\ref{it:seqend}), $r<\alpha(G_w)$, and any $\ell$ with $\ell\le q$ suffices by an analog argument.
The above construction defines the sequence $C^u_0,\ldots,C^u_{p^u}$. We will now prove that it is an SU-sequence.

As observed above, $C^u_0=C^v_0\cup C^w_0=(I\cap V_v)\cup (I\cap V_w)=I\cap V_u$, so Property~\ref{it:begin} is satisfied. 
Since $|C_{b+1}|>|C_{b}|$ and $|C_{c+1}|>|C_c|$ holds for any $b$ and $c$ (Property~\ref{it:increase}), it follows that $|C_{a+1}|>|C_a|$ holds for any $a<p^u$, which proves Property~\ref{it:increase} for the new sequence.

Now we prove Property~\ref{it:newverts}. Consider $a'\le a$ with $C^u_a=C^v_b\cup C^w_c$ and $C^u_{a'}=C^v_{b'}\cup C^w_{c'}$. So $b'\le b$ and $c'\le c$.
Assume w.l.o.g. that $C^u_{a+1}=C^v_{b+1}\cup C^w_c$.
Then we can write
\[
(C^u_{a+1}\bs C^u_a)\cap C^u_{a'}\subseteq (C^v_{b+1}\bs C^v_b)\cap (C^v_{b'}\cup C^w_{c'})=\emptyset.
\]
For the last equality, we used
\begin{itemize}
 \item Property~\ref{it:newverts} for $v$ to conclude that $(C^v_{b+1}\bs C^v_b)\cap C^v_{b'}=\emptyset$, and 
 \item the observations that $C^v_{b+1}\subseteq V_v$, $C^w_{c'}\subseteq V_w$, and $V_v\cap V_w=\emptyset$ to conclude that $(C^v_{b+1}\bs C^v_b)\cap C^w_{c'}=\emptyset$.
\end{itemize}

It remains to prove Property~\ref{it:shortseq} for the new sequence. First note that $\RIS_0^I(u)=\alpha(G_u)=\alpha(G_v)+\alpha(G_w)$, so we may end the sequence when $q=\alpha(G_v)$ and $r=\alpha(G_w)$.
Now consider any index $a<p^u$, such that we constructed $C^u_{a+1}$ from $C^u_a=C^v_b\cup C^w_c$ using the above method. 
We will prove for all $\ell'\in \{0,\ldots,|I\cap V_u|\}$ with
$|C_a|<\RIS^I_{\ell'}(u)$ that there exists an $\ell'$-\TARseq\ in $G_u$ from $C_a$ to $C_{a+1}$ that only adds tokens on $C_{a+1}\bs C_a$.

First, we show that $|C^u_a|<\RIS^I_{\ell'}(u)$ implies that $\RIS^I_{\ell'-r}(v)>q$ or $\RIS^I_{\ell'-q}(w)>r$. 
Consider an $\ell'$-\TARseq\ $I_0,\ldots,I_q$ from $I\cap V_u$ to an \indset\ $J$ of $G_u$ with $|J|=\RIS^I_{\ell'}(u)>|C^u_a|=q+r$. Let $j$ be the first index such that $|I_j\cap V_v|\ge q+1$ or $|I_j\cap V_w|\ge r+1$. Clearly, such an index $j$ exists, and $j\ge 1$ holds since $|C^u_a|\ge |C^u_0|$ by Property~\ref{it:increase}.
W.l.o.g. assume that $|I_j\cap V_v|\ge q+1$. Then define $I'_i=I_i\cap V_v$ for all $i\in \{0,\ldots,j\}$. By choice of $i$, for all $i\in \{0,\ldots,j\}$ it holds that 
$|I_i\cap V_w|\le r$, and therefore $|I'_i|\ge \ell'-r$.
So the sequence $I'_0,\ldots,I'_j$ is an $(\ell'-r)$-\TARseq\ for $G_v$ from $I\cap V_v$ to an \indset\ $I'_j$ with $|I'_j|\ge q+1$, and thus $\RIS_{\ell'-r}^I(v)\ge q+1$.

From this fact we conclude that for any $\ell'\in \{0,\ldots,|I\cap V_u|\}$ with $|C_a|<\RIS^I_{\ell'}(u)$, it holds that $\ell'\le \ell$, where $\ell$ is the value chosen in~(\ref{it:ellchoice}).
We conclude the proof of Property~\ref{it:shortseq} by showing that there exists an $\ell$-\TARseq\ in $G_u$ from $C^u_a$ to $C^u_{a+1}$ that only adds tokens on $C^u_{a+1}\bs C^u_a$ (which is then obviously also an $\ell'$-\TARseq). 

Consider the case that we have chosen $C^u_{a+1}=C^v_{b+1}\cup C^w_c$.
Then $\RIS^I_{\ell-r}(v)>|C^v_b|$, so by using Property~\ref{it:shortseq} for the SU-sequence for $v$, there exists an $(\ell-r)$-\TARseq\ in $G_v$ from $C^v_b$ to $C^v_{b+1}$ that only adds tokens on $C^v_{b+1}\bs C^v_b$.
If we apply the same token additions to $C^u_a=C^v_b\cup C^w_c$, then this yields the desired $\ell$-\TARseq\ from $C^u_a$ to $C^u_{a+1}$, since any \indset\ in this sequence contains  $r$ vertices of $V_w$. If $C^u_{a+1}=C^v_{b}\cup C^w_{c+1}$, then $\RIS^I_{\ell-q}(w)>|C^w_c|$, and the proof is analog.

Summarizing, we have now shown that for the constructed sequence $C^u_0,\ldots,C^u_{p^u}$, all properties from Definition~\ref{def:SUseq} hold, and therefore it is an SU-sequence for $u$, based on $I$, which concludes the proof of the lemma.
\QED

A straightforward induction proof based on Proposition~\ref{propo:SUseq_leaf}, Lemma~\ref{lem:SUseq_join} and Lemma~\ref{lem:SUseq_union} now yields the following statement.
\begin{thm}
\label{thm:SUseqsExist}
Let $T$ be a cotree of a graph $G$, and let $I$ be an \indset\ of $G$. For every node $u\in V(T)$, there exists an SU-sequence based on $I$.
\end{thm}

Combined with Proposition~\ref{propo:SU_ends_with_maxindset} and Lemma~\ref{lem:SUseq_usage}, this shows that for any value of $k$ such that there exists a $k$-\TARseq\ in $G$ from $I$ to some maximum \indset\ of $G$, then there exists a short $k$-\TARseq\ of this type.

\begin{thm}
\label{thm:ShortTARseqToMax}
Let $G$ be a graph on $n$ vertices, let $T$ be a cotree of $G$ with root $r$, let $I$ be an \indset\ of $G$, and let $k$ be an integer such that $\RIS^I_k(r)=\alpha(G)$. 
Then there exists a $k$-\TARseq\ from $I$ to some maximum independent set $J$ of $G$ with length at most $2n-|I|-\alpha(G)$.
\end{thm}

\PF
By Theorem~\ref{thm:SUseqsExist}, there exists an SU-sequence $C_0,\ldots,C_p$ for the root node $r$. 
Since $\RIS^I_k(r)=\alpha(G)=|C_p|$ (Proposition~\ref{propo:SU_ends_with_maxindset}),  Lemma~\ref{lem:SUseq_usage} shows that there exists a $k$-\TARseq\ from $I$ to $C_p$ of length $2(|\bigcup_{j=0}^{p} C_j|)-|C_0|-|C_{p}|\le 2n-|I|-\alpha(G)$.
\QED

Theorem~\ref{thm:ShortTARseqToMax} can be used to prove the existence of a linear length \TARseq\ between any two \indset s $A$ and $B$ with $A\tar_k^G B$, by reconfiguring both to a common reachable maximum \indset. 
There are however two problems with this approach: first, even though $A\tar_k^G B$ holds, it may be that $A$ and $B$ cannot reach any maximum \indset\ of $G$. This is remedied by considering an appropriate subgraph $G'$ of $G$, such that $A\tar_k^{G'} B$ holds, and both $A$ and $B$ can reach a maximum \indset\ of $G'$. Lemma~\ref{lem:ReachableSubgraph} below indicates how this graph $G'$ can be chosen -- it suffices to simply omit all vertices that are not in any \indset\ that can be reached from $A$ or $B$. 
Secondly, even if both $A$ and $B$ can both reach a maximum \indset\ of a graph $G$ (i.e. $\RIS^A_k(r)=\alpha(G)=\RIS^B_k(r)$), it may be that $G$ has multiple maximum \indset s, and Theorem~\ref{thm:ShortTARseqToMax} does not specify which one is reachable. In fact, from the construction of the SU-sequences it can be seen that different choices of $A$ and $B$ may lead to different maximum \indset s. Therefore, to conclude the proof, we also need to demonstrate that short \TARseq s exist between any pair of maximum \indset s that can reach each other. This is done in the next lemma.

\begin{lem}
\label{lem:ShortTARbetweenMaxIndSets} 
Let $A$ and $B$ be two maximum \indset s of a cograph $G$. If $A\tar_k^G B$, then there exists a $k$-\TARseq\ from $A$ to $B$ of length $|A\Delta B|$.
\end{lem}

\PF
We prove the statement by induction over $|A\Delta B|$.
Let $T$ be a cotree of $G$.
If $A=B$ then there is nothing to prove, so assume now that $|A\Delta B|\ge 1$.
Define a {\em difference node} to be a node $u\in V(T)$ with 
$A\cap V_u=\emptyset$ and $B\cap V_u\not=\emptyset$ or with
$A\cap V_u\not=\emptyset$ and $B\cap V_u=\emptyset$.

Consider a {\em join node $u$ with children $v$ and $w$ such that $u$ is not a difference node, but $v$ and $w$ are}. We first argue that such a node exists.
Since $A\bs B\not=\emptyset$, there exists at least one difference node (a leaf of $T$). Considering the root $r$, there exists also at least one node that is not a difference node. So we may consider a difference node $v$ for which the parent $u$ is not a difference node.
W.l.o.g. assume that $A\cap V_v\not=\emptyset$ and $B\cap V_v=\emptyset$. 
Since $u$ is not a difference node, $B\cap V_w\not=\emptyset$, where $w$ is the other child of $u$. If $u$ is a union node, then we can add $A\cap V_v$ to $B$, such that the result is a larger \indset\ (since $V_u$ is a module with $V_u\cap B\not=\emptyset$, and vertices in $A\cap V_v$ are not adjacent to vertices in $B\cap V_w$), a contradiction with the maximality of $B$. So $v$ is a join node. Since $A$ is an \indset, it follows that $A\cap V_w=\emptyset$, and therefore $w$ is also a difference node, which proves that a node $u$ with the stated properties exists.

Next, we prove that $|A\bs V_u|\ge k$ and $|B\bs V_u|\ge k$.
Consider a $k$-\TARseq\ $I_0,\ldots,I_p$ from $A$ to $B$.
By choice of $u$, this sequence contains an \indset\ that contains no vertices of $V_u$. Let $I_i$ be the first such \indset\ in the sequence. So $i\ge 1$ and $I_{i-1}=I_{i}\cup \{x\}$ for some $x\in V_u$. Because $I_{i-1}$ is an \indset\ and $V_u$ is a module, $I_{i}$ contains no vertices that are adjacent to any vertex in $V_u$. So $(A\cap V_u)\cup I_i$ is an \indset, which implies that $|A|=\alpha(G)\ge |A\cap V_u|+|I_i|\ge |A\cap V_u|+k$, and thus $|A\bs V_u|\ge k$. Analogously, $|B\bs V_u|\ge k$ follows.

Since $V_u$ is a module of $G$, it follows that $(A\bs V_u)\cup (B\cap V_u)$ and $(B\bs V_u)\cup (A\cap V_u)$ are also \indset s. 
In fact, since their cardinalities sum to $2\alpha(G)$, and neither set can be larger than $\alpha(G)$, it follows that both are maximum \indset s of $G$.
Denote $A'=(A\bs V_u)\cup (B\cap V_u)$.

Since $|A\bs V_u|\ge k$, a $k$-\TARseq\ from $A$ to $A'$ can be obtained by first removing all tokens from $A\cap V_u$, and next adding tokens on all of $B\cap V_u$. This sequence has length $|A\Delta A'|$. By induction, there is a $k$-\TARseq\ from $A'$ to $B$ of length $|A'\Delta B|$. Because $|A\Delta A'|+|A'\Delta B|=|A\Delta B|$, this proves the statement.
\QED

\begin{lem}
\label{lem:ReachableSubgraph}
Let $T$ be a cotree for $G$, 
and let $I$ be an \indset\ of $G$ such that for all $v\in V(G)$, there exists an \indset\ $J$ with $I\tar_k^G J$ and $v\in J$. Then $\RIS_k^I=\alpha(G)$.
\end{lem}

\PF
Let $I^*$ be a maximum \indset\ of $G$. 
By induction over the cotree $T$, we will prove that Claim~A below holds for every node $u\in V(T)$. Applying Claim~A for to the root node of $T$ proves the lemma statement.

\medskip
\noindent
{\em Claim A:} There exists an \indset\ $J$ of $G$ with $I^*\cap V_u\subseteq J\cap V_u$ and $I\tar_k^G J$.

\medskip
Suppose $u\in V(T)$ is a (trivial) leaf node. Then Claim~A follows immediately from the assumption.

\medskip
Suppose $u$ is a join node, with children $v$ and $w$.
We may assume w.l.o.g. that $I^*\cap V_w=\emptyset$. 
By induction, there exists an \indset\ $J$ with $I\tar_k^G J$ and $I^*\cap V_v\subseteq J\cap V_v$. 
Therefore, $I^*\cap V_u= I^*\cap V_v \subseteq J\cap V_v \subseteq J\cap V_u$, which proves Claim~A for $u$.

\medskip
Finally, suppose $u$ is a union node, with children $v$ and $w$.
If $I^*\cap V_u=\emptyset$ then Claim~A follows trivially for $u$, so assume this is not the case. Since $I^*$ is now a maximum \indset\ of $G$ with $I^*\cap V_u\not=\emptyset$, and $V_u$ is a module of $G$ that is the disjoint union of $V_v$ and $V_w$, it follows that $I^*\cap V_v$ and $I^*\cap V_w$ are maximum \indset s for $G_v$ and $G_w$, respectively. Indeed, if this would not be the case, then the size of $I^*$ can be increased by replacing $I^*\cap V_v$ or $I^*\cap V_w$ by arbitrary maximum \indset s of $G_v$ and $G_w$ respectively, while maintaining an \indset, a contradiction.

By induction, there exists an \indset\ $J_v$ with $I\tar_k^G J_v$ and $I^*\cap V_v\subseteq J_v\cap V_v$, and there exists an \indset\ $J_w$ with $I\tar_k^G J_w$ and $I^*\cap V_w\subseteq J_w\cap V_w$. It follows that $J_v\cap V_v$ and $J_w\cap V_w$ are maximum \indset s of $G_v$ and $G_w$ respectively.
We may now apply (module) Lemma~\ref{lem:moduleB} (with $V_u$, $V_v$, $V_w$, $I$, $J_v$ and $J_w$ in the roles of $M$, $M_1$, $M_2$, $A$, $B_1$, $B_2$, respectively), to conclude that there exists an \indset\ $J$ of $G$ with $I\tar_k^G J$, $J\cap V_v=J_v\cap V_v$, and $J\cap V_w=J_w\cap V_w$. So $I^*\cap V_u=(I^*\cap V_v)\cup (I^*\cap V_w)\subseteq J\cap V_u$. 
This proves Claim~A for $u$.\QED

Now we can prove the main theorem from this section.

\begin{thm}
\label{thm:diameter}
Let $G$ be a cograph on $n$ vertices, with \indset s $A$ and $B$ such that $A\tar_k^G B$. Then there exists a $k$-\TARseq\ from $A$ to $B$ of length at most $4n-|A|-|B|$.
\end{thm}

\PF
For an \indset\ $I$ of $G$, call a vertex $v\in V(G)$ {\em $k$-accessible from $I$} if there exists an \indset\ $J$ with $I\tar_k^G J$ and $v\in J$. Since $A\tar_k^G B$, and $\tar_k^G$ is an equivalence relation, it follows that for every vertex $v\in V(G)$, $v$ is $k$-accessible from $A$ if and only if it is $k$-accessible from $B$. 
So we may consider the subgraph $G'$ induced by all vertices that are $k$-accessible from $A$. For any \indset\ $J$ of $G$ it now holds that $A\tar_k^G J$ if and only if $J\subseteq V(G')$ and $A\tar_k^{G'} J$, and the same statement holds if we replace $A$ by $B$. 

Since $G'$ is an induced subgraph of $G$, it is again a cograph, so we may choose $T$ to be a cotree of $G'$, with root $r$. Denote $n'=|V(G')|$.
By definition, $G'$ satisfies the conditions of Lemma~\ref{lem:ReachableSubgraph}, for both $I=A$ and $I=B$, so 
$\RIS^A_k(r)=\alpha(G')=\RIS^B_k(r)$. 
Theorem~\ref{thm:ShortTARseqToMax} then shows that there exist $k$-\TARseq s from $A$ and $B$ to maximum \indset s $A'$ and $B'$ of $G'$ respectively, of length at most $2n'-|A|-|A'|$ and $2n'-|B|-|B'|$.
Lemma~\ref{lem:ShortTARbetweenMaxIndSets} shows that there exists a $k$-\TARseq\ from $A'$ to $B'$ of length $|A'\Delta B'|$.
Combining these three $k$-\TARseq s gives a $k$-\TARseq\ from $A$ to $B$ in $G'$ of length at most $4n'-|A|-|B|-|A'|-|B'|+|A'\Delta B'|\le 4n'-|A|-|B|\le 4n-|A|-|B|$. Since $G'$ is an induced subgraph of $G$, this is also a $k$-\TARseq\ for $G$.
\QED

This immediately yields:

\begin{corol}
\label{corol:TARdiameter}
For any cograph $G$ on $n$ vertices and integer $k$, components of $\TAR_k(G)$ have diameter at most $4n-2k$.
\end{corol}

Combining the previous corollary with Lemma~\ref{lem:TJisTAR} yields:

\begin{corol}
\label{corol:TJdiameter}
For any cograph $G$ on $n$ vertices and integer $k$, components of $\TJ_k(G)$ have diameter at most $2n-k$.
\end{corol}

\section{Discussion}
\label{sec:discussion}

In this paper, we showed that the TAR-Reachability problem (and thus the TJ-Reachability problem) can be solved efficiently for any graph that admits a cograph decomposition into  graphs that satisfy certain properties (Theorem~\ref{thm:main_alg}) -- call this a {\em good graph class}. 
Chordal graphs are given as an example of a good graph class. In fact, this might be generalized to even-hole-free graphs, provided that the following question can be answered affirmatively: can $\alpha(G)$ be computed in polynomial time if $G$ is an even-hole-free graph? This is a well-known open question~\cite{Vus10}, and also a negative answer (i.e. NP-hardness proof) would be interesting (see~\cite{KMM12}).

Another good graph class is the class of claw-free graphs, which will be shown in another paper~\cite{BKW}. Finally, Theorem~\ref{thm:main_alg} easily applies to any graph class such that graphs on $n$ vertices admit a cograph decomposition into $O(\log n)$ sized graphs: in this case, a trivial (exponential time) exhaustive search procedure can be applied to the base graphs, such that the total complexity is still polynomial in $n$.

Together, this shows that the TAR-Reachability problem can be solved efficiently for quite a rich graph class. Considering the fact that TAR-Reachability is PSPACE-hard for perfect graphs~\cite{KMM12}, the boundary between hard and easy graph classes for this problem starts to become clear.

Recall that cographs are exactly the graphs of cliquewidth two, and of modular width two~\cite{CO00}. Generalizing our result to an efficient algorithm for graphs of bounded cliquewidth may be too challenging; a more reasonable goal is to first consider graphs of bounded modular width. The {\em modular width} of a graph is the largest number of vertices of a prime graph appearing at some node of its unique modular decomposition tree~\cite{CO00,GLO13}. Is there a polynomial time algorithm for TAR-Reachability for all graphs of modular width at most $k$, for every constant $k$?

\medskip
The following two questions related to independent set reconfiguration in cographs are still open: first, what is the complexity of deciding whether there exists a $k$-\TARseq\ of length at most $\ell$ between two \indset s of a cograph? (Recall that for general graphs, this is strongly NP-hard~\cite{KMM12}.) Secondly, what is the complexity of deciding whether $\TAR_k(G)$ is connected, if $G$ is a cograph? We expect that a variant of our DP algorithm can be used to show that this problem can be decided in polynomial time. 

\medskip 

\bibliographystyle{plain}

\begin{thebibliography}{10}

\bibitem{Bod93tourist}
H.~L. Bodlaender.
\newblock A tourist guide through treewidth.
\newblock {\em Acta Cybernetica}, 11:1--23, 1993.

\bibitem{PSPR_FSTTCS12}
P.~Bonsma.
\newblock Rerouting shortest paths in planar graphs.
\newblock In {\em FSTTCS 2012}, volume~18 of {\em LIPIcs}, pages 337--349.
  Schloss Dagstuhl - Leibniz-Zentrum fuer Informatik, 2012.

\bibitem{TCS13}
P.~Bonsma.
\newblock The complexity of rerouting shortest paths.
\newblock {\em Theoretical Computer Science}, 510:1 -- 12, 2013.

\bibitem{BC09}
P.~Bonsma and L.~Cereceda.
\newblock Finding paths between graph colourings: {PSPACE}-completeness and
  superpolynomial distances.
\newblock {\em Theoretical Computer Science}, 410(50):5215--5226, 2009.

\bibitem{BKW}
P.~Bonsma, M.~Kami\'{n}ski, and M.~Wrochna.
\newblock Independent set reconfiguration in claw-free graphs.
\newblock working paper, 2014.

\bibitem{CHJ08}
L.~Cereceda, J.~van~den Heuvel, and M.~Johnson.
\newblock Connectedness of the graph of vertex-colourings.
\newblock {\em Discrete Applied Mathematics}, 308(5--6):913--919, 2008.

\bibitem{CHJ09}
L.~Cereceda, J.~van~den Heuvel, and M.~Johnson.
\newblock Mixing 3-colourings in bipartite graphs.
\newblock {\em European Journal of Combinatorics}, 30(7):1593--1606, 2009.

\bibitem{CHJ11}
L.~Cereceda, J.~van~den Heuvel, and M.~Johnson.
\newblock Finding paths between 3-colorings.
\newblock {\em Journal of Graph Theory}, 67(1):69--82, 2011.

\bibitem{CPS85}
D.~Corneil, Y.~Perl, and L.~Stewart.
\newblock A linear recognition algorithm for cographs.
\newblock {\em SIAM Journal on Computing}, 14(4):926--934, 1985.

\bibitem{CMR00}
B.~Courcelle, J.A. Makowsky, and U.~Rotics.
\newblock Linear time solvable optimization problems on graphs of bounded
  clique-width.
\newblock {\em Theory of Computing Systems}, 33(2):125--150, 2000.

\bibitem{CO00}
B.~Courcelle and S.~Olariu.
\newblock Upper bounds to the clique width of graphs.
\newblock {\em Discrete Applied Mathematics}, 101(1–3):77 -- 114, 2000.

\bibitem{EW12}
C.E.J. Eggermont and G.J. Woeginger.
\newblock {Motion planning with pulley, rope, and baskets}.
\newblock In {\em STACS 2012}, volume~14 of {\em LIPIcs}, pages 374--383.
  Schloss Dagstuhl--Leibniz-Zentrum fuer Informatik, 2012.

\bibitem{GLO13}
J.~Gajarsk{\'y}, M.~Lampis, and S.~Ordyniak.
\newblock Parameterized algorithms for modular-width.
\newblock In {\em IPEC 2013}, volume 8246 of {\em LNCS}, pages 163--176.
  Springer, 2013.

\bibitem{Gav72}
F.~Gavril.
\newblock Algorithms for minimum coloring, maximum clique, minimum covering by
  cliques, and maximum independent set of a chordal graph.
\newblock {\em SIAM Journal on Computing}, 1(2):180--187, 1972.

\bibitem{GKM09}
P.~Gopalan, P.G. Kolaitis, E.~Maneva, and C.H. Papadimitriou.
\newblock The connectivity of boolean satisfiability: Computational and
  structural dichotomies.
\newblock {\em SIAM Journal on Computing}, 38(6), 2009.

\bibitem{HD05}
R.A. Hearn and E.D. Demaine.
\newblock {PSPACE}-completeness of sliding-block puzzles and other problems
  through the nondeterministic constraint logic model of computation.
\newblock {\em Theoretical Computer Science}, 343(1--2):72--96, 2005.

\bibitem{JvdH13}
J.~van~den Heuvel.
\newblock The complexity of change.
\newblock {\em Surveys in Combinatorics 2013}, pages 127--160, 2013.

\bibitem{ID11}
T.~Ito and E.D. Demaine.
\newblock Approximability of the subset sum reconfiguration problem.
\newblock In {\em TAMC 2011}, volume 6648 of {\em LNCS}, pages 58--69.
  Springer, 2011.

\bibitem{IDH11}
T.~Ito, E.D. Demaine, N.J.A. Harvey, C.H. Papadimitriou, M.~Sideri, R.~Uehara,
  and Y.~Uno.
\newblock On the complexity of reconfiguration problems.
\newblock {\em Theoretical Computer Science}, 412(12--14):1054--1065, 2011.

\bibitem{KMM11TCS}
M.~Kami\'{n}ski, P.~Medvedev, and M.~Milani\v{c}.
\newblock Shortest paths between shortest paths.
\newblock {\em Theoretical Computer Science}, 412(39):5205--5210, 2011.

\bibitem{KMM12}
M.~Kami\'{n}ski, P.~Medvedev, and M.~Milani\v{c}.
\newblock Complexity of independent set reconfigurability problems.
\newblock {\em Theoretical Computer Science}, 439:9--15, 2012.

\bibitem{MNRSS13}
A.E. Mouawad, N.~Nishimura, V.~Raman, N.~Simjour, and A.~Suzuki.
\newblock On the parameterized complexity of reconfiguration problems.
\newblock In {\em IPEC 2013}, volume 8246 of {\em LNCS}, pages 281--294.
  Springer, 2013.

\bibitem{Schrijver}
A.~Schrijver.
\newblock {\em Combinatorial Optimization: Polyhedra and Efficiency}, volume~24
  of {\em Algorithms and Combinatorics}.
\newblock Springer, Berlin, 2003.

\bibitem{Vus10}
K.~Vu{\v{s}}kovi{\'c}.
\newblock Even-hole-free graphs: a survey.
\newblock {\em Applicable Analysis and Discrete Mathematics}, 4(2):219--240,
  2010.

\end{thebibliography}

\end{document}